\PassOptionsToPackage{table}{xcolor}

\documentclass[preprint]{aastex63}

\usepackage[frozencache=true,cachedir=.]{minted} 
\usepackage{physics}
\usepackage{xspace}
\usepackage{tabstackengine}
\usepackage{xparse}

\usepackage{siunitx}
\stackMath

\newenvironment{nalign}{
    \begin{equation}
    \begin{aligned}
}{
    \end{aligned}
    \end{equation}
    \ignorespacesafterend
}

\def \snewpy {SNEWPY\xspace}
\def \snowglobes {SNOwGLoBES\xspace}
\def \sntools {sntools\xspace}

\received{\today}
\revised{TBD}
\accepted{TBD}
\submitjournal{ApJ}

\shorttitle{SNEWPY}
\shortauthors{Baxter et al. (SNEWS 2.0 Collaboration)}

\begin{document}

\title{\snewpy: A Data Pipeline from Supernova Simulations to Neutrino Signals}

\correspondingauthor{James P. Kneller}
\email{jim\_kneller@ncsu.edu}

\author{Amanda L. Baxter}
\affiliation{Department of Physics and Astronomy, Purdue University, West Lafayette, IN, USA}

\author[0000-0001-5537-4710]{Segev BenZvi}
\affiliation{University of Rochester, Rochester, NY, USA}

\author{Joahan Castaneda Jaimes}
\affiliation{Caltech, Pasadena, CA, USA}

\author{Alexis Coleiro}
\affiliation{Université de Paris, CNRS, AstroParticule et Cosmologie, Paris, France}

\author[0000-0003-1801-8121]{Marta Colomer Molla}
\affiliation{Inter-university Institute for High Energies, Université libre de Bruxelles, Brussels, Belgium}

\author{Damien Dornic}
\affiliation{Aix Marseille Univ, CNRS/IN2P3, CPPM, Marseille, France}

\author{Tomer Goldhagen}
\affiliation{University of North Carolina - Chapel Hill, Chapel Hill, NC, USA}

\author{Anne Graf}
\affiliation{NC State University, Raleigh, NC, USA}

\author[0000-0002-7321-7513]{Spencer Griswold}
\affiliation{University of Rochester, Rochester, NY, USA}

\author[0000-0002-1018-9383]{Alec Habig}
\affiliation{University of Minnesota Duluth, Duluth, MN, USA}

\author{Remington Hill}
\affiliation{Laurentian University, Sudbury, ON, Canada}

\author[0000-0001-6142-6556]{Shunsaku Horiuchi}
\affiliation{Virginia Tech, Blacksburg, VA 24061, USA}
\affiliation{Kavli IPMU (WPI), UTIAS, The University of Tokyo, Kashiwa, Chiba 277-8583, Japan}

\author[0000-0002-3502-3830]{James P. Kneller}
\affiliation{NC State University, Raleigh, NC, USA}

\author[0000-0001-7594-2746]{Rafael F. Lang}
\affiliation{Department of Physics and Astronomy, Purdue University, West Lafayette, IN, USA}

\author[0000-0002-1460-3369]{Massimiliano Lincetto}
\affiliation{Astronomisches Institut, Ruhr-Universitaet Bochum, 44780 Bochum, Germany}

\author[0000-0002-5350-8049]{Jost Migenda}
\affiliation{King's College London, London, UK}

\author[0000-0002-8734-2147]{Ko Nakamura}
\affiliation{Fukuoka University, Fukuoka, Japan}

\author[0000-0002-8228-796X]{Evan O'Connor}
\affiliation{The Oskar Klein Centre, Department of Astronomy, Stockholm University, AlbaNova, SE-106 91 Stockholm, Sweden}

\author[0000-0003-2913-8057]{Andrew Renshaw}
\affiliation{University of Houston, Houston, TX, USA}

\author[0000-0002-7007-2021]{Kate Scholberg}
\affiliation{Duke University, Durham, NC, USA}

\author{Navya Uberoi}
\affiliation{University of Rochester, Rochester, NY, USA}

\author{Arkin Worlikar}
\affiliation{Georgia Tech, Atlanta, GA, USA}
\collaboration{999}{(The SNEWS Collaboration)}

\begin{abstract}
Current neutrino detectors will observe hundreds to thousands of neutrinos from a Galactic supernovae, and future detectors will increase this yield by an order of magnitude or more. With such a data set comes the potential for a huge increase in our understanding of the explosions of massive stars, nuclear physics under extreme conditions, and the properties of the neutrino. However, there is currently a large gap between supernova simulations and the corresponding signals in neutrino detectors, which will make any comparison between theory and observation very difficult. \snewpy is an open-source software package which bridges this gap. The \snewpy code can interface with supernova simulation data to generate from the model either a time series of neutrino spectral fluences at Earth, or the total time-integrated spectral fluence. Data from several hundred simulations of core-collapse, thermonuclear, and pair-instability supernovae is included in the package. This output may then be used by an event generator such as \sntools or an event rate calculator such as \snowglobes. Additional routines in the \snewpy package automate the processing of the generated data through the \snowglobes software and collate its output into the observable channels of each detector. In this paper we describe the contents of the package, the physics behind \snewpy, the organization of the code, and provide examples of how to make use of its capabilities.
\end{abstract}



\section{Introduction}

The neutrino signal from a supernova in the Milky Way Galaxy will be a golden opportunity to advance our understanding of how stars explode and probe the properties of the neutrino.
For reviews of the physics potential of a supernova neutrino signal we refer the reader to several recent reviews \citep{2012ARNPS..62...81S,Mirizzi:2015eza,2016ARNPS..66..341J,2018JPhG...45d3002H}.
Many different detectors worldwide will record events from the supernova and naturally there is much interest from experimenters about what they might observe.
Present-day detectors, such as Super-Kamiokande \citep{Ikeda:2007sa}, Borexino \citep{Monzani:2006jg}, KamLAND \citep{Eguchi:2002dm}, LVD \citep{Agafonova:2007hn,2007APh....27..254A}, or the dedicated supernova burst detector HALO \citep{2008JPhCS.136d2077D}, are expected to record tens to hundreds of thousands of neutrino events each from a Galactic or nearby core-collapse supernova (CCSN).
Such a neutrino burst would also be recorded with very high statistics in detectors such as IceCube \citep{Halzen:1994xe,Halzen:1995ex,2011A&A...535A.109A} and KM3NeT \citep{2021EPJC...81..445A} but without event-by-event reconstruction.
Ever larger and more ambitious detectors are under construction or proposed, including DUNE \citep{Acciarri:2016crz,Acciarri:2015uup,Acciarri:2016ooe}, Hyper-Kamiokande \citep{Hyper-Kamiokande:2018ofw}, and JUNO \citep{An:2015jdp}.
The physical and astronomical potential of megaton scale detectors has been considered \citep{Suzuki:2001rb,2011PhRvD..83l3008K}.
In addition, dark matter detectors are becoming of sufficient size that they, too, are capable of detecting a significant number of neutrinos from a CCSN burst via coherent elastic neutrino-nucleus scattering \citep{Horowitz:2003cz, Lang:2016zhv, Lai:2021uxd}.
Other types of supernovae or compact object mergers involving neutron stars also produce neutrino bursts that are detectable if they occur sufficiently nearby \citep{Odrzywolek2011a,2016PhRvD..94b5026W,2017PhRvD..95d3006W,PhysRevD.96.103008,2003MNRAS.342..673R,2009PhRvD..80l3004C,Lin:2019piz}.

With so many detectors capable of detecting the next Galactic or near-Galactic supernova, there is a need for theoretical predictions of neutrino emission from supernova simulations and the resulting neutrino signals in detectors on Earth.
To determine how well these detectors can provide quantitative information about the supernova explosion mechanism and nuclear physics under extreme conditions, we need a suite of neutrino signal templates to compare observations with theory. However, there is presently a substantial gulf between the data from a supernova simulation and the neutrino spectra at Earth.
This gap has been bridged on a few occasions \citep{Kneller:2008PhRvD..77d5023K,2011PhRvD..84h5023R,PhysRevD.91.065016,2009PhRvL.103g1101G}---see also \cite{2016PhRvD..94b5026W,2017PhRvD..95d3006W,PhysRevD.96.103008}---but there are not enough such data sets to cover the wide range of available supernova models and they are time consuming to generate.

The \snewpy code has been written with the intent of bridging this gap.
\snewpy provides a consistent interface to hundreds of supernova simulation data sets to extract the neutrino emission.
It can then convolve this with a prescription for the flavor transformation to generate the spectral fluence (time-integrated flux) reaching a detector on Earth either in a set of time bins or over the entire simulation time window.
This output can be processed with an event generator such as \sntools \citep{Migenda:2021hnl} or an event rate calculator such as \snowglobes \citep{SNOwGLoBES}. \snewpy also includes routines that automate the processing of the data with \snowglobes and collate its output to determine the total event rates in the observable channels of each detector.

In this paper we give an overview of the \snewpy package, its capabilities, and the large library of simulation data sets. The overall organization of the code is described in section~\ref{sec:structure} before proceeding to discuss in detail \snewpy's interfaces to simulation data in section~\ref{sec:models}, the flavor transformations prescriptions in section~\ref{sec:transforms}, and the \snowglobes interface in section~\ref{sec:snowglobes}. To demonstrate the capabilities of the software and how it is used in practice, in section~\ref{sec:examples} we provide two examples of applications of \snewpy: the first example is a complete data pipeline using \snewpy's interface with \snowglobes, and the second example shows how it can be integrated into the event generator \sntools. 


\section{The Structure of SNEWPY}\label{sec:structure}

\begin{figure}
    \includegraphics[width=\textwidth]{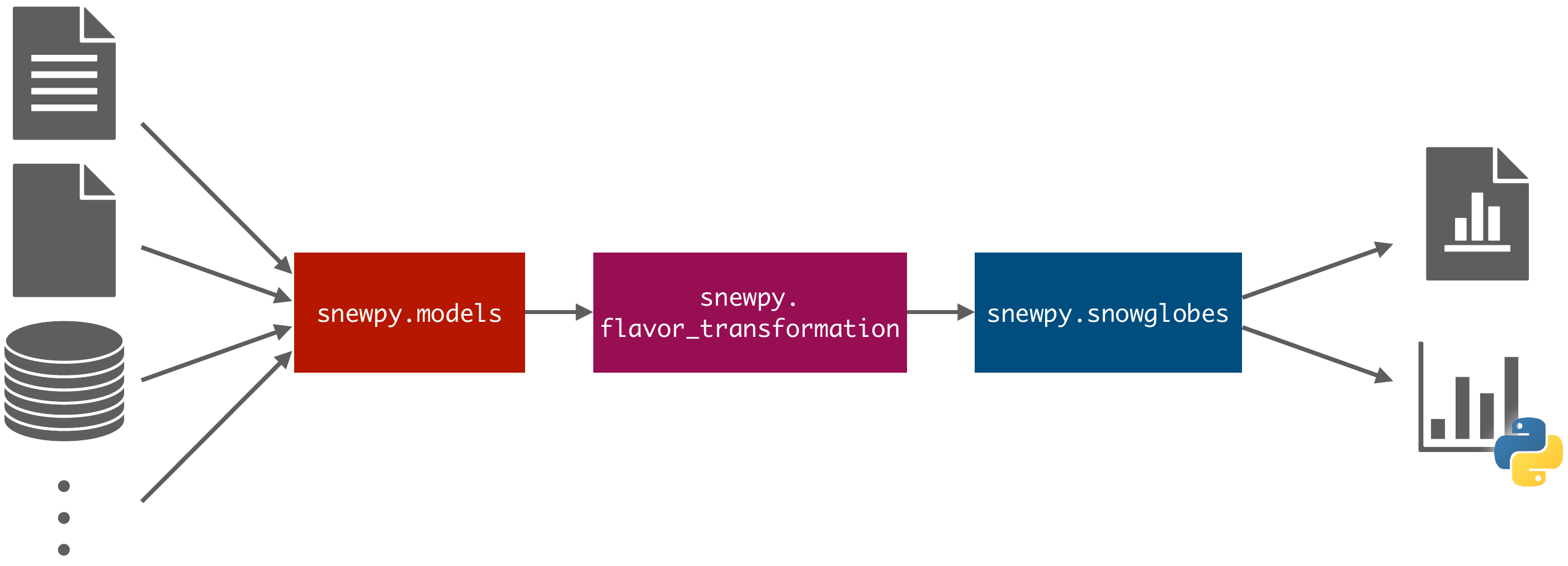}
\caption{Flowchart showing the complete \snewpy pipeline. \snewpy supports a wide variety of input formats and can output results as plots or as a Python dictionary for further analysis.}
\label{fig:Flowchart}
\end{figure}

\snewpy is an open-source package written in Python, designed to bridge the gap between supernova simulations and detector observations.
It is built upon NumPy \citep{harris2020array} and SciPy \citep{Virtanen:2019joe} and uses Astropy \citep{Astropy:2013muo, Price-Whelan:2018hus} for model I/O and unit conversions.
Figure~\ref{fig:Flowchart} shows the high-level structure of a complete supernova data pipeline implemented in \snewpy. This pipeline takes an input file from a supernova simulation, then
\begin{itemize}
\setlength{\itemsep}{-0.25em}
\item extracts neutrino fluxes produced in the supernova as a function of time, energy, and flavor,
\item applies a flavor transformation prescription to determine the fluxes reaching Earth, and
\item runs them through the detector response software \snowglobes,
\end{itemize}
before providing the computed event rates per detector and interaction channel either in the form of figures or as structured data for further processing.

Matching this set of steps, \snewpy is divided into three main modules\footnote{Notation: \texttt{monospace} font is used to refer to modules and file names, \textbf{bold} font is used for class names, and \textsl{italic} font indicates class members.}: \texttt{snewpy.models}, which interfaces with the simulation data that comes with \snewpy, \texttt{snewpy.flavor\_transformation}, which implements the different flavor transformation prescriptions, and \texttt{snewpy.snowglobes}, which integrates with \snowglobes and provides functions to generate \snowglobes-formatted data files, runs them through \snowglobes for a chosen set of neutrino detectors and finally collates the resulting outputs. \snewpy was designed in this way so that the user can insert or extract data at the interfaces between the components. For example, the user may have an alternative method (such as an analytic formula) for generating the neutrino spectra at Earth and therefore does not need to generate a time series from a simulation. Another user could use \snewpy to provide a consistent interface to different supernova models or a large library of flavor transformations, without having to run the \snowglobes software.
In what follows, we describe these three main modules of \snewpy in more detail.


\section{Supernova Neutrino Models}\label{sec:models}

Core-collapse supernova models depend on many different factors---both physical parameters of the progenitor (e.\,g. its mass and metallicity) and implementation parameters of the simulation (e.\,g. the degree of spatial symmetry and the equation of state). The choice of parameters strongly affects the predicted neutrino emission and thus what we detect at Earth. To best estimate the sensitivity of a detector (or set of detectors) to a wide variety of explosion models and parameter values, it is common practice to use a broad sample of simulation models. It is better still if those models are provided by different modeling groups using different numerical algorithms and approaches. Unfortunately, this often means simulation data is provided in different formats. Rather than attempt to reduce all simulations to a common format, \snewpy solves this problem by providing an extensible module called \texttt{models} which contains an abstract \textbf{SupernovaModel} base class with the absolute minimal functionality to undertake the calculation of the neutrino spectra at Earth from the simulation. The \textbf{SupernovaModel} class defines just three member functions:
\begin{itemize}
    \item \textsl{get\_time}: access the list of snapshot times from the simulation;
    \item \textsl{get\_initial\_spectra}: obtain the neutrino spectra as a function of time, energy, flavor, and angle at the surface of emission (neutrinosphere) within the progenitor;
    \item \textsl{get\_transformed\_spectra}: obtain neutrino spectra as a function of time, energy, and flavor after some flavor transformation (see Section \ref{sec:transforms}).
\end{itemize}

\begin{figure}
    \includegraphics[width=\textwidth]{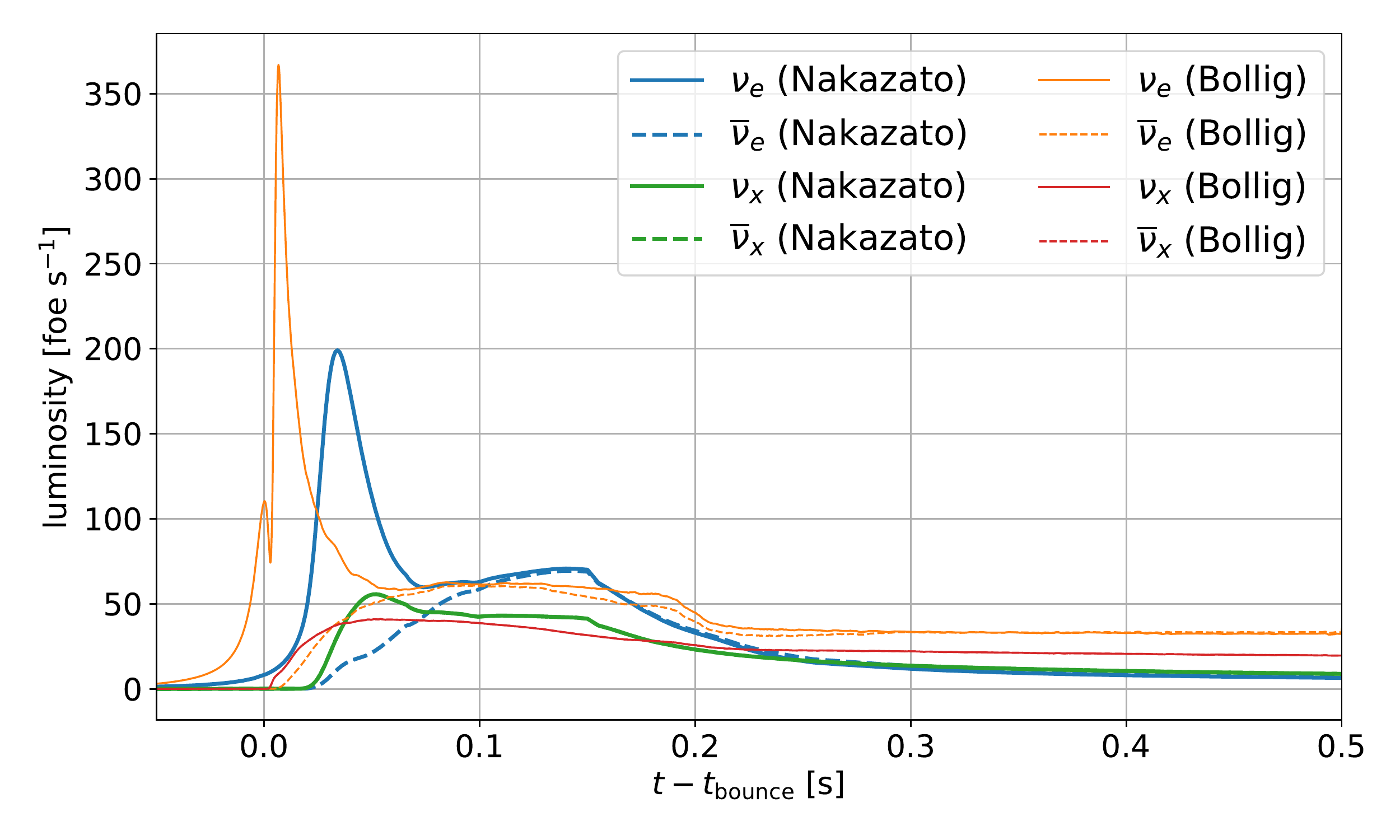}
\caption{Luminosity of different neutrino flavors as a function of time for the \textbf{nakazato-shen-z0.004-trev100ms-s20.0} model (thick lines, blue and green) and the \textbf{Bollig\_2016/s27.0c} model (thin lines, orange and red). Both models come with \snewpy and were originally presented in \cite{Nakazato_2013} and \cite{Mirizzi:2015eza}, respectively.}
\label{fig:luminosity-comparison}
\end{figure}

Specific supernova models are implemented as subclasses of \textbf{SupernovaModel}. Each subclass must contain custom implementations of \textsl{get\_time} and \textsl{get\_initial\_spectra}, since these depend on the format of the input files. At the present time, \texttt{snewpy.models} contains twelve \textbf{SupernovaModel} subclasses, which are named according to the modeling group or publication describing the simulations. A list of available models and their properties is provided in Table~\ref{tab:models} and two example models are shown in figure~\ref{fig:luminosity-comparison}.

\begin{table}[ht]
\centering
\footnotesize
\begin{tabular}{l >{\raggedright\arraybackslash}p{2.25cm} >{\raggedright\arraybackslash}p{2.95cm} >{\raggedright\arraybackslash}p{8.5cm}}
{\bf Model} & {\bf Masses [M$_\odot$]} & {\bf Time Range [s]} & {\bf Comment}
\\ \hline
{\bf Nakazato\_2013} & 13, 20, 30, 50 & $-0.05$ - $20.0$ (SN) $-0.14$ - $0.84$ (BH) & Binned spectra from \cite{Nakazato_2013} reparameterized in terms of $\langle E\rangle$ and $\alpha$. Uses Shen and LS220 equations of state (EOS) and includes a black hole (BH) formation scenario.
\\ \hline
{\bf Tamborra\_2014} & 20, 27 & $\phantom{+}0.006$ - $0.338$ (20) $\phantom{+}0.011$ - $0.552$ (27) & 3D models from \cite{2014PhRvD..90d5032T}, using emission direction with maximum SASI signal.
\\ \hline
{\bf OConnor\_2015} & 40 & $-0.378$ - $0.537$ & BH-forming simulation \citep{2015ApJS..219...24O} using a $40~M_{\odot}$ progenitor from \cite{2007PhR...442..269W} and LS220 EOS.
\\ \hline
{\bf Sukhbold\_2015} & 9.6, 27 & $-0.35$ - $15.44$ & PROMETHEUS-VERTEX simulation data presented in \cite{2002AA...396..361R} and \cite{2016ApJ...821...38S}, with LS220 and SFHo EOS.
\\ \hline
{\bf Bollig\_2016} & 11.2, 27 & $-0.17$ - $7.60$ (11) $-0.34$ - $7.60$ (27) & The s11.2c and s27.0c models shown in Fig. 17 of \cite{2016NCimR..39....1M}, using the LS220 EOS.
\\ \hline
{\bf Walk\_2018} & 15 & $\phantom{+}0.01$ - $0.33$ & Models from \cite{2018PhRvD..98l3001W} demonstrating effect of progenitor rotation on SASI oscillations.
\\ \hline
{\bf Walk\_2019} & 40 & $\phantom{+}0.01$ - $0.57$ & BH-forming simulation from \cite{2020PhRvD.101l3013W} with strong SASI features prior to black hole formation.
\\ \hline
{\bf Fornax\_2019} & 9, 10, 12, 13, 14, 15, 16, 19, 25, 60 & $\phantom{+}0.01$ - $1.04$ (9) $\phantom{+}0.01$ - $0.40$ (60) & Full 3D simulation of \cite{2019MNRAS.482..351V} produced using the FORNAX code \citep{2019ApJS..241....7S}.
\\ \hline
{\bf Warren\_2020} & 9.0 - 100.0 (200 in total) & $-0.22$ - $4.7$, depending on the simulation & 1D FLASH simulations with STIR presented in \cite{2020ApJ...898..139W}.
\\ \hline
{\bf Kuroda\_2020} & 20 & $\phantom{+}0.00$ - $0.476$, depending on rotation and magnetic field & 3D simulation of a magnetized rotating star from \citet{Kuroda:2020pta} using a \SI{20}{M_\odot} progenitor by \citet{2007PhR...442..269W} and SFHo EOS.
\\ \hline
{\bf Fornax\_2021} & 12, 13, 14, 15, 16, 17, 18, 19, 20, 21, 22, 23, 25, 26, 27 & $-0.21$ - $4.49$ (12) $-0.31$ - $4.59$ (27) &  Axisymmetric models simulated to 4.5~s post-bounce using the FORNAX code, described in \cite{2021Natur.589...29B}.
\\ \hline
{\bf Zha\_2021} & 16, 17, 18, 19, 19.89, 20, 21, 22.39, 23, 24, 25, 26, 30, 33 & $-0.21$ - $2.02$ (16) $-0.30$ - $0.35$ (33) &  Failing CCSN simulations with a hybrid EOS including a hadron-quark phase transition \citep{Zha:2021fbi}.
\end{tabular}
\normalsize
\caption{CCSN simulation models with neutrino emission tables included in \snewpy. User-defined CCSN models can be created by subclassing {\bf SupernovaModel} in the module {\tt snewpy.models}.}
\label{tab:models}
\end{table}

Across these models, neutrino fluxes from hundreds of core-collapse simulations are available and can be downloaded by \snewpy on demand.
In addition, \snewpy has data from two pair-instability supernovae (PISNe) from \cite{PhysRevD.96.103008}, and two Type-Ia supernovae \citep{2016PhRvD..94b5026W,PhysRevD.96.103008}.%
\footnote{Note that the PISNe and Type-Ia data sets differ from the core-collapse simulations in that (a) flavor transformation have already been included, (b) for the Type Ia data sets there are multiple lines of sight, and (c) the data is already in \snowglobes format.}
Given so many models we do not attempt to describe them all here and refer the reader to their associated literature. 

Each \textbf{SupernovaModel} subclass is able to construct a neutrino spectrum $\Phi_{\alpha}$ at the neutrinosphere for the neutrino species $\nu_e$, $\nu_x$, $\bar{\nu}_e$, and $\bar{\nu}_x$. Here, $x$ stands for the $\mu$ and $\tau$ flavors, which are treated as identical in almost all current supernova simulations, including those in table~\ref{tab:models}.%
\footnote{Many current simulations go further and do not distinguish between $\nu_x$ and $\bar{\nu}_x$, so there are just three distinct spectra.}
Once simulations that treat all six flavors---3 neutrino plus 3 antineutrino flavors---as separate become more frequent, we expect to update \snewpy to accommodate this.

The customized \textsl{get\_initial\_spectra} member function of each model class extracts the neutrino spectra $\Phi(E_\nu)$ from the simulation data if it is provided but, more commonly, the method constructs a spectrum according to a parameterization. A common parameterization of this spectrum---though not required by \snewpy---is in terms of the luminosity $L_\nu$, the mean energy $\expval{E_\nu}$, and the spectral shape parameter $\alpha$ \citep{Keil:2002in, Tamborra:2012ac}:

\begin{equation}
\Phi(E_\nu) = \frac{L_\nu}{\expval{E_\nu}} \,\frac { (\alpha + 1)^{\alpha + 1} }{ \expval{E_\nu}\,\Gamma (\alpha + 1) } \,\qty(\frac{E_\nu}{\expval{E_\nu}})^{\alpha} \exp \qty(- \frac{(\alpha + 1)\,E_\nu }{\expval{E_\nu}}) \label{eq:Phi}. 
\end{equation}
This shape parameter can be calculated from the energy moments $\langle E_{\nu}^k \rangle$ via

\begin{equation}
\frac{\expval{E_\nu^k}}{\expval{E_\nu^{k-1}}}
= \frac{k+\alpha}{1+\alpha}\expval{E_\nu}.
\end{equation}
If we use $k=2$, the most common additional moment output from a simulation, we find

\begin{equation}
\alpha = \frac{2\expval{E_\nu}^2 - \expval{E_\nu^2}}{\expval{E_\nu^2}-\expval{E_\nu}^2}.
\end{equation}
Whatever the method used to construct the initial spectra, the initial spectra generally evolve $\Phi_{\alpha}$ with time as the simulation proceeds. 

The last method of the \textbf{SupernovaModel} class is \textsl{get\_transformed\_spectra}. This method convolves the initial spectra with a flavor transformation prescription, which we now describe. 


\section{The Flavor Transformation Prescriptions}
\label{sec:transforms}

The second component of \snewpy is the flavor transformation prescriptions which relate the neutrino fluxes produced in the supernova to those arriving on Earth. This library of prescriptions in \snewpy accounts for effects of propagation through the outer layers of the star, neutrino decay during vacuum propagation to Earth, or mixing with sterile neutrinos. The user may easily create new prescriptions to include additional scenarios or vary oscillation parameters for one of the existing prescriptions to test the sensitivity of a detector.
In this section we first describe the general form of these transformations and then their implementation in \snewpy.

\subsection{General Form of Flavor Transformations}

The spectral fluxes of the three neutrino flavors at Earth can be arranged into a column vector, $F_{F}(r_{\oplus}) = \left( F_e(r_{\oplus}), F_{\mu}(r_{\oplus}), F_{\tau}(r_{\oplus}) \right)^{T}$, where we use the subscript $F$ to indicate the neutrino flavor basis, 
and $r_{\oplus}$ indicates the location of Earth. 
When we refer to a generic element of the flavor basis, we shall use Greek subscripts $\alpha$ and $\beta$; the subscripts $e$, $\mu$ and $\tau$ are the specific flavor of the neutrinos, while the subscript $x$ indicates either $\mu$ or $\tau$.
The neutrino spectra at Earth are not those that were emitted at the neutrinosphere inside the progenitor, but have been mixed on their way to Earth by flavor transformation effects. 
The flavor transformation of neutrinos is a quantum mechanical phenomenon that occurs due to the mismatch between the flavor states of the neutrinos and the eigenstates of the free Hamiltonian. For a complete discussion of the phenomenon we refer the reader to the reviews mentioned previously \citep{Mirizzi:2015eza,2018JPhG...45d3002H}.
Flavor transformation of the neutrinos occurs while they are within the supernova and then at some point on their trip to Earth the neutrinos decohere and arrive in their mass eigenstates.
We shall denote quantities which are in the mass basis by the subscript $M$ and when we wish to refer to a generic mass states we shall use the italic subscripts $i$, $j$; the subscripts 1, 2 and 3 are specific mass states. 

The spectral flux of a particular flavor at Earth is the incoherent sum of the parts of the spectral flux of each mass state with a given flavor, which can be written as 

\begin{equation}
F_{F}(r_{\oplus}) = D\, F_{M}(r_{\oplus}). \label{eq:F}
\end{equation}

The elements of the matrix $D$ are given by $D_{\alpha i} = \norm{U_{V,\alpha i} }^2$, where $U_{V,\alpha i}$ are the elements of the vacuum mixing matrix for neutrinos. The antineutrino fluxes in the flavor basis are related to the antineutrino fluxes in the mass basis by the same matrix $D$. We shall denote antineutrino quantities by an overbar. The relationship for antineutrinos is thus $\bar{F}_{F}(r_{\oplus}) = D \bar{F}_{M}(r_{\oplus})$.
The column vector of the spectral fluxes of the mass states are the diagonal elements of the spectral flux matrix $\mathcal{F}_{M}$, constructed by

\begin{equation}
\mathcal{F}_{M} = c \int \rho_M \, \cos\theta \,d\Omega 
\end{equation}
where $c$ is the speed of light, $\rho_M$ is the density matrix in the mass basis, $\theta$ is the angle with respect to the line from the supernova to Earth, and the integral is over all the neutrino propagation angles.
The conversion of the flux matrix to a column vector---and the ignoring of the off-diagonal elements of the flux matrix---is the effect of decoherence.
Accounting for decoherence, $F_{M}$ can be written as

\begin{equation}
F_{M}(r_{\oplus}) = \sum_{j} \ket{\nu_j} \mel{\nu_j}{\mathcal{F}_{M}}{\nu_j}
\end{equation}
where the $\qty{\ket{\nu_j}}$ are basis vectors in the mass basis. The same conversion occurs for the antineutrinos. 
The matrix $\mathcal{F}_M$ is 

\begin{equation}
\mathcal{F}_{M}(r_{\oplus}) = \frac{1}{4\pi d^2} \,\Phi_{M}(r_{\oplus})
\end{equation}
where $d$ is the distance of the supernova from Earth, and $\Phi_{M}(r_{\oplus})$ is the spectral number luminosity matrix in the mass basis at Earth. 
The matrix $\Phi_{M}$ at Earth is related to the matter basis spectral matrix at the neutrinosphere $\Phi_{M}(R_{\nu})$ - the matter basis becomes the mass basis in the vacuum - by 

\begin{equation}
\Phi_{M}(r_{\oplus})= S_{M}(r_{\oplus},R_{\nu}) \,\Phi_{M}(R_{\nu})\, S_{M}^{\dagger}(r_{\oplus},R_{\nu})
\end{equation}
with $S_{M}(r_{\oplus},R_{\nu})$ being the propagator of the matter/mass states and $R_{\nu}$ indicating the neutrinosphere. Note that this formula does not account for absorption, emission, or scattering of neutrinos. Finally, the matrix $\Phi_{M}(R_{\nu})$ is related to the spectral matrix in the flavor basis at the neutrinosphere by 

\begin{equation}
\Phi_{M}(R_{\nu})= U^{\dagger}\, \Phi_{F}(R_{\nu})\, U
\end{equation}
with $U$ as the so-called matter mixing matrix at the neutrinosphere. 
The matrix $U$, and the equivalent matrix for the antineutrinos $\bar{U}$, depend upon the mass ordering and the initial density. The spectral matrix at the neutrinosphere in the flavor basis is taken to be pure diagonal: 

\begin{equation}
\Phi_{F}(R_{\nu}) = \left( \begin{array}{ccc} \Phi_e(R_{\nu}) & 0 & 0 \\ 0 & \Phi_{\mu}(R_{\nu}) & 0 \\ 0 & 0 & \Phi_{\tau}(R_{\nu}) \end{array} \right) = \sum_{\beta} \ketbra{\nu_\beta} \Phi_{\beta}(R_{\nu}) \label{eq:PhiRnu}
\end{equation}
where the $\qty{\ket{\nu_\beta}}$ are the basis vectors in the flavor basis.
Note that in practice  the spectra $\Phi_{\beta}(R_{\nu})$ are usually taken to be the spectra at the largest radius in the simulation and not the actual neutrinosphere. 
Putting all this together so as to relate $F_{F}(r_{\oplus}) $ to $\Phi_{F}(R_{\nu})$ and $\bar{F}_{F}(r_{\oplus})$ to ${\bar{\Phi}}_{F}(R_{\nu})$, we obtain

\begin{eqnarray}
F_{F}(r_{\oplus}) &= &\frac{1}{4\pi d^2}\,  D \sum_{i} \sum_{\beta} \Phi_{\beta}(R_{\nu}) \ketbra{\nu_i} \, S_{M}(r_{\oplus},R_{\nu})\, U^{\dagger}\, \ketbra{\nu_\beta}\, U\, S_{M}^{\dagger}(r_{\oplus},R_{\nu}) \, \ket{\nu_i}  \label{eq:FatEarth}\\
\bar{F}_{F}(r_{\oplus}) &= &\frac{1}{4\pi d^2}\,  D \sum_{i}  \sum_{\beta} {\bar{\Phi}}_{\beta}(R_{\nu}) \ketbra{{\bar{\nu}}_i} \, {\bar{S}}_{M}(r_{\oplus},R_{\nu})\, \bar{U}^{\dagger}\,\ketbra{{\bar{\nu}}_\beta}\, \bar{U}\, {\bar{S}}_m^{\dagger}(r_{\oplus},R_{\nu}) \, \ket{{\bar{\nu}}_i}.
\label{eq:FbaratEarth}
\end{eqnarray}

In these equations, the quantities $\mel{\nu_i}{S_{M}(r_{\oplus},R_{\nu})\, U^{\dagger}}{\nu_\beta}$ have the physical interpretation of being the probability amplitudes that a neutrino emitted as a particular flavor $\beta$ reaches Earth in mass state $i$. The propagators $S_{M}$ and $\bar{S}_{M}$ and the mixing matrices $U$ and $\bar{U}$ are the quantities which can vary depending upon the evolution of the matter states through the mantle of the supernova, and the flavor transformation scenario.

Equations (\ref{eq:FatEarth}) and (\ref{eq:FbaratEarth}) can be simplified considerably by summarizing the effect of transformations as 

\begin{align}
F_{e} & = \frac{1}{4\pi d^2} \left[\, p_{ee} \Phi_{\nu_e} + p_{ex} \Phi_{\nu_x} \right] \label{eq:Fe} \\
F_{x} & = \frac{1}{4\pi d^2} \left[\,p_{xe} \Phi_{\nu_e} + p_{xx} \Phi_{\nu_x} \right] \\
{\bar{F}}_e & = \frac{1}{4\pi d^2} \left[\,\bar{p}_{ee} \bar{\Phi}_{\nu_e} + \bar{p}_{ex} \bar{\Phi}_{\nu_x} \right]\\
{\bar{F}}_x & = \frac{1}{4\pi d^2} \left[\, \bar{p}_{xe} \bar{\Phi}_{\nu_e} + \bar{p}_{xx} \bar{\Phi}_{\nu_x} \right] \label{eq:Fxbar}.
\end{align}
where the quantities $\qty{p_{\alpha\beta}}$ and $\qty{{\bar{p}}_{\alpha\beta}}$ are the various survival and transition probabilities that a neutrino emitted as a particular flavor $\beta$ is detected as flavor $\alpha$ at Earth. Equations (\ref{eq:Fe}) to (\ref{eq:Fxbar}) are generalizations of those found in \cite{2000PhRvD..62c3007D,2018JPhG...45d3002H} and elsewhere, since we shall consider cases where $p_{\alpha\beta}\neq p_{\beta\alpha}$. 

\subsection{Implementation of Flavor Transformations in \snewpy} \label{sec:transforms-implementation}

\snewpy implements neutrino flavor transformations in its \texttt{flavor\_transformation} module. This module contains an abstract base class \textbf{FlavorTransformation} which defines a minimal interface for computing neutrino survival and transformation probabilities used in equations~\eqref{eq:Fe} to \eqref{eq:Fxbar}. For example, the survival probability $p_{ee}=\Pr(\nu_e\to\nu_e)$ is computed using the member function \textsl{prob\_ee}, and the transition probability $p_{ex}=\Pr(\nu_x\to\nu_e)$ is computed using the member function \textsl{prob\_ex}. For forward compatibility, \snewpy has been written such that the transition probabilities can depend on time and neutrino energy.

Using this interface, it is easy to support many different kinds of flavor transformation scenarios with subclasses that inherit the interface of \textbf{FlavorTransformation}. Currently, the module \texttt{snewpy.flavor\_transformation} supports fifteen transformation scenarios: six for the normal mass ordering (NMO), six for the inverted mass ordering (IMO), and three which are independent of the mass ordering. The list of transformations included in \snewpy is provided in Table~\ref{tab:xforms}.
The values used for the neutrino mixing parameters are contained inside two instances of a class called \textbf{MixingParameters} that is also part of \snewpy, one instance for the NMO and another for the IMO. They are set by default to the values from \cite{2020JHEP...09..178E} but may be modified by the user.

\begin{table}[ht]
\centering
\footnotesize
\begin{tabular}{lc >{\raggedright\arraybackslash}p{10.7cm}}
{\bf Prescription Name} & {\bf Hierarchy} & {\bf Comment}
\\ \hline
{\bf NoTransformation} & --- & No flavor transformation applied.
\\ \hline
{\bf CompleteExchange} & --- & Electron flavors completely swapped with a heavy lepton flavor.
\\ \hline
{\bf AdiabaticMSW} & NMO, IMO & Adiabatic neutrino evolution for normal or inverted hierarchy.
\\ \hline
{\bf NonAdiabaticMSWH} & NMO, IMO & The H resonance is nonadiabatic and the L resonance is adiabatic.
\\ \hline
{\bf TwoFlavorDecoherence} & NMO, IMO & 50\% mixing between whichever states mix at the H resonance.
\\ \hline
{\bf ThreeFlavorDecoherence} & --- & 33\% mixing between all flavors and both neutrinos and antineutrinos.
\\ \hline
{\bf NeutrinoDecay} & NMO, IMO & Adiabatic evolution through the supernova mantle followed by decay of the heaviest neutrino mass state to the lightest during vacuum propagation to Earth. Uses the approximation that the energy of the neutrino does not change.
\\ \hline
{\bf AdiabaticMSWes} & NMO, IMO & Mixing for four neutrino flavors, where the fourth mass state is the heaviest and the new ``es'' MSW resonance is adiabatic. The mass ordering refers to the three lightest neutrinos.
\\ \hline
{\bf NonAdiabaticMSWes} & NMO, IMO & Mixing for four neutrino flavors, where the fourth mass state is the heaviest and the new ``es'' MSW resonance is nonadiabatic. The mass ordering refers to the three lightest neutrinos.
\end{tabular}
\normalsize
\caption{Flavor transformation models included in \snewpy. Explanations and probabilities for each transformation prescription are provided in Appendix~\ref{sec:appendix2}. User-defined transformations can be created by subclassing {\bf FlavorTransformation} in the module {\tt snewpy.flavor\_transformation}.}
\label{tab:xforms}
\end{table}

\begin{figure}
    \includegraphics[width=\textwidth]{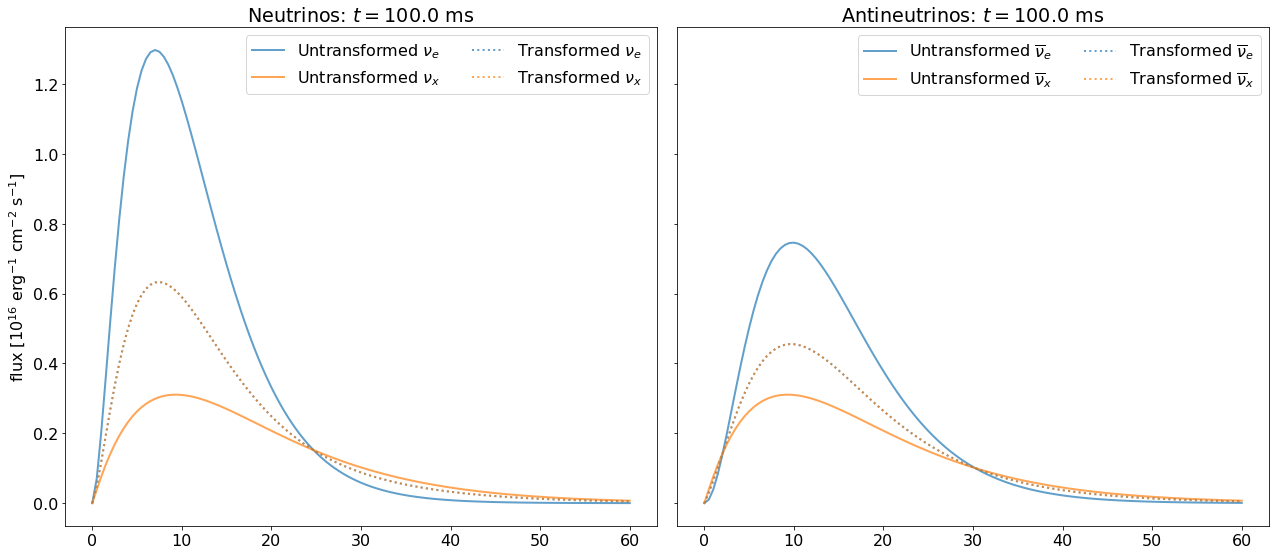}
\caption{The initial (solid lines) and transformed (dashed lines) spectral flux for neutrinos (left panels) and antineutrinos (right panels) using the \textbf{ThreeFlavorDecoherence} prescription and a distance to the supernova of 10 kpc. Both mass orderings are equivalent in this prescription. We show the \textbf{nakazato-shen-z0.004-trev100ms-s20.0} model \citep{Nakazato_2013} that comes with \snewpy, at the simulation time of \SI{100}{ms} and at a supernova distance of \SI{10}{kpc}.}
\label{fig:ThreeFlavorDecoherence}
\end{figure}
With both the model and flavor transformation prescriptions now defined, the previously mentioned \textsl{get\_transformed\_spectra} member function of each model class is seen to be the implementation of equations (\ref{eq:Fe}) to (\ref{eq:Fxbar}). For the purposes of illustration, we show in figure \ref{fig:ThreeFlavorDecoherence} an example of how flavor transformations affect the neutrino spectra. This example uses the \textbf{ThreeFlavorDecoherence} prescription, which mixes the initial spectra for the neutrinos so that the flux of each flavor at Earth is equal (and the same for antineutrinos).


\section{\snewpy's Interface with \snowglobes}\label{sec:snowglobes}

The module \texttt{snewpy.snowglobes} contains functions to interact with the \snowglobes software. This interaction typically occurs in three steps: generating the data files of the neutrino fluences at Earth in the \snowglobes format, processing those files through the \snowglobes software, and then collating the data output from \snowglobes.

\subsection{Generating \snowglobes Input Files}

The \texttt{snewpy.snowglobes} module contains the two functions \textsl{generate\_time\_series} and \textsl{generate\_fluence}. The \textsl{generate\_time\_series} function constructs a set of data files at a set of snapshot times from a chosen simulation. For each snapshot, the code extracts (or constructs) the neutrino spectra at the neutrinosphere. It then applies a user-selected flavor transformation prescription to those spectra and scales with a user-selected supernova distance to generate the flux at Earth.
Finally, the function writes the fluence---the flux multiplied by the time bin width---for the  snapshot to a \snowglobes-formatted file and collates the files for different snapshot times into a single compressed output file. A number of options are available to the user to control how many snapshots are created and their spacing in time.

The \textsl{generate\_fluence} function works similarly but applies a time integration of the fluence in each time bin to compute the total fluence. 

\subsection{Simulating Detector Effects with \snowglobes}

The \texttt{snewpy.snowglobes} module contains a function named \textsl{simulate} which takes a compressed file, such as those generated by \textsl{generate\_time\_series} or \textsl{generate\_fluence}, and runs the contents through the \snowglobes software. The bulk of the \textsl{simulate} function is a translation into Python of the script \texttt{supernova.pl} that comes with \snowglobes: \textsl{simulate} also creates a \texttt{supernova.glb} file then invokes \texttt{supernova}, the executable in \snowglobes. 
The mandatory arguments passed to \textsl{simulate} are the location of the \snowglobes installation directory and the path to the compressed input file. Another, optional argument is the name of a detector. If this argument is not supplied, \snowglobes is repeatedly invoked for all supported detectors. There is no output from the \textsl{simulate} function itself; instead, upon successful execution of the function a large set of files (each containing the event rate in a given set of energy bins for a particular interaction channel due to a particular flavor) is generated in the \snowglobes output folder. We refer the reader to the \snowglobes documentation \citep{SNOwGLoBES} for details about the interaction channels that \snowglobes computes and how the calculation is done.

\subsection{Collating Information from \snowglobes}
Finally, the \texttt{snewpy.snowglobes} module contains a function named \textsl{collate}, which collates the output generated by \snowglobes into the observable channels of each detector in the four combinations of weighted and unweighted event rates and with and without applying detector energy smearing.

Like the \textsl{simulate} function, the mandatory arguments passed to \textsl{collate} are the location of the \snowglobes installation directory and the name of the compressed input file. Again, an optional final argument is the name of the detector that was chosen for the \snowglobes event rate calculation and if this argument is not supplied, it is assumed all detectors in the \snowglobes suite were used. The \textsl{collate} function writes the collated data to a new compressed file, whose name is based on the name of the file that was run through \snowglobes, and returns a dictionary of the collated results. Additional optional arguments to the function will cause it to generate simple Matplotlib \citep{Hunter:2007} histogram figures from the collated data, which will be also placed in the compressed file, or to delete \snowglobes output files.


\section{\snewpy Usage Examples}\label{sec:examples}

Included in the \snewpy repository are many Jupyter notebooks containing many instances of using the software for various purposes. Here we provide two short examples. 

\subsection{Using \snewpy with \snowglobes}\label{sec:snowglobes-example}

In this section we provide a simple example of using \snewpy with the \snowglobes software. 
This particular script reproduces the Super-Kamiokande entry in Table 4 of  \cite{2021NJPh...23c1201A} for the \SI{11.2}{M_\odot} model assuming adiabatic MSW oscillations with the normal mass ordering (4045 neutrinos).  
A script that calculates the entire Table 4 in \cite{2021NJPh...23c1201A} is one of the previously mentioned Jupyter notebooks that are included in the package.

\begin{minted}{python}
from snewpy import snowglobes

SNOwGLoBES_path = "/path/to/snowglobes/"  # directory where SNOwGLoBES is located
SNEWPY_model_dir = "/path/to/snewpy/models/"  # directory with model input files

distance = 10  # Supernova distance in kpc
detector = "wc100kt30prct"  # Name of SNOwGLoBES detector model to use
modeltype = 'Bollig_2016'  # Model type from snewpy.models
model = 's11.2c'  # Name of model
transformation = 'AdiabaticMSW'  # Desired flavor transformation

# Construct file system path of model file and name of output file
model_path = SNEWPY_model_dir + "/" + modeltype + "/" + model
outfile = modeltype + "_" + model + "_" + transformation

# Now, do the main work:
print("Generating fluence files ...")
tarredfile = snowglobes.generate_fluence(model_path, modeltype, transformation,
                                         distance, outfile)

print("Simulating detector effects with SNOwGLoBES ...")
snowglobes.simulate(SNOwGLoBES_path, tarredfile, detector_input=detector)

print("Collating results ...")
tables = snowglobes.collate(SNOwGLoBES_path, tarredfile, detector_input=detector,
                            skip_plots=True)

# Use results to print the number of events in different interaction channels
key = f"Collated_{outfile}_{detector}_events_smeared_weighted.dat"
total_events=0
for i, channel in enumerate(tables[key]['header'].split()):
    if i == 0:
        continue
    # Scale to Super-K inner volume (32 kt)
    n_events = 0.32 * sum(tables[key]['data'][i])
    total_events += n_events
    print(f"{channel:10}: {n_events:.3f} events")

print("Total events in Super-K-like detector:", total_events)
\end{minted}


\subsection{Using \snewpy as Part of the Event Generator \sntools}\label{sec:sntools}
\sntools \citep{Migenda:2021hnl} is an event generator that takes neutrino fluxes from supernova simulations and generates a list of resulting neutrino interactions in a detector with a realistic distribution of event time as well as energy and direction of outgoing particles.
This list can then be used with a full detector simulation and event reconstruction toolchain for situations in which the approximate treatment of detector effects in \snowglobes is insufficient.
In a recent update, \sntools integrated \snewpy to benefit from the large number of supernova models and flavor transformations it implements.

For example, generating a set of neutrino events in the Hyper-Kamiokande detector \citep{Hyper-Kamiokande:2018ofw} for the first \SI{500}{ms} of one of the models included in \snewpy at a supernova distance of \SI{50}{kpc} can be done as follows:
\begin{minted}{bash}
# install sntools (this automatically installs SNEWPY as a dependency)
pip install sntools
# download supernova model files that are part of SNEWPY
python -c 'import snewpy; snewpy.get_models("Bollig_2016")'
# run sntools using an input file from SNEWPY
sntools SNEWPY_models/Bollig_2016/s27.0c --format SNEWPY-Bollig_2016 --distance 50
        --detector HyperK --starttime 0 --endtime 500
\end{minted}

\snewpy{}’s modular design also makes it possible to use its flavor transformations with unsupported input fluxes. The following example shows how to apply \snewpy{}’s \textbf{ThreeFlavorDecoherence} flavor transformation to an input file format that is natively supported by \sntools but not by \snewpy.\footnote{This particular file is from the simulations by \citet{Nakazato_2013}. Note: Unlike the \textbf{Nakazato\_2013} class in \texttt{snewpy.models}, which uses reparameterized and reformatted files, sntools natively supports the original file format from \url{http://asphwww.ph.noda.tus.ac.jp/snn/index.html}.}
\begin{minted}{bash}
# download sample input file
curl https://raw.githubusercontent.com/JostMigenda/sntools/v1.0b2/fluxes/intp2001.data
     -o intp2001.data
# run sntools using a flavor transformation from SNEWPY
sntools intp2001.data --format nakazato --detector HyperK --distance 50
        --starttime 0 --endtime 500 --transformation SNEWPY-ThreeFlavorDecoherence
\end{minted}

At the moment, \sntools does not support all supernova models and flavor transformations listed in sections \ref{sec:models} and \ref{sec:transforms-implementation}, respectively, since some require additional physical parameters such as sterile neutrino mixing angles. Improvements are expected in future versions of \sntools.


\section{Summary}\label{sec:summary}

The \snewpy software package connects supernova simulations with detector response software such as \snowglobes and \sntools, allowing users to calculate the expected event rates in various neutrino detectors for each model. We expect \snewpy will prove useful to modelers and theorists interested in what detectors will observe given some new piece of physics in a simulation, and to experimentalists wishing to evaluate the sensitivity of their detector to supernova neutrinos. In the future we plan to enhance the capabilities of \snewpy and suggestions from the community about the features we should add are warmly welcome. Modelers interested in adding their simulations to the model library are encouraged to contact us. 


\begin{acknowledgments}

This work is supported by the National Science Foundation “Windows on the Universe: the Era of Multi-Messenger Astrophysics” Program: “WoU-MMA: Collaborative Research: A Next-Generation SuperNova Early Warning System for Multimessenger Astronomy” through Grant Nos. 1914448, 1914409, 1914447, 1914418, 1914410, 1914416, and 1914426.
This work is also supported at NC State by DOE grant DE-FG02-02ER41216, at King’s College London by STFC, and at Stockholm University by the Swedish Research Council (Project No. 2020-00452).

\end{acknowledgments}

\software{Astropy \citep{Astropy:2013muo, Price-Whelan:2018hus}, Matplotlib \citep{Hunter:2007}, NumPy \citep{harris2020array}, SciPy \citep{Virtanen:2019joe}, SNOwGLoBES \citep{SNOwGLoBES}, sntools \citep{Migenda:2021hnl}}

\appendix

\section{Derivation of the probabilities}
\label{sec:appendix2}
In this appendix we provide the equations for the survival and transition probabilities that appear in equations (\ref{eq:Fe}) to (\ref{eq:Fxbar}) for all 15 flavor transformation prescriptions that \snewpy currently implements.

\subsection{The Extreme Cases}

The first two transformations implemented in \snewpy are \textbf{NoTransformation} and \textbf{CompleteExchange}. For \textbf{NoTransformation}, the set of probabilities is given by

\begin{nalign}
    p_{ee} &= 1, & 
    p_{ex} &= 0
    \\
    p_{xx} &= 1, & 
    p_{xe} &= 0
    \\
    & \\
    {\bar{p}}_{ee} &= 1, &
    {\bar{p}}_{ex} &= 0
    \\
    {\bar{p}}_{xx} &= 1, & 
    {\bar{p}}_{xe} &= 0,
\end{nalign}
while \textbf{CompleteExchange} corresponds to
\begin{nalign}
    p_{ee} &= 0, & 
    p_{ex} &= 1
    \\
    p_{xx} &= 0.5, & 
    p_{xe} &= 0.5
    \\
    & \\
    {\bar{p}}_{ee} &= 0, &
    {\bar{p}}_{ex} &= 1
    \\
    {\bar{p}}_{xx} &= 0.5, & 
    {\bar{p}}_{xe} &= 0.5
\end{nalign}

\subsection{The Three Flavor Mixing Prescriptions}

For nontrivial cases of mixing between three active flavors, we only need to compute the elements of the $D$ matrix from the ``electron'' flavor row in terms of the vacuum mixing angles $\theta_{12}$ and $\theta_{13}$. The expressions in the `electron' flavor row of the matrix are:

\begin{align} 
D_{e1} &= \cos^{2}\theta_{12}\,\cos^{2}\theta_{13}, \\
D_{e2} &= \sin^{2}\theta_{12}\,\cos^{2}\theta_{13}, \\
D_{e3} &= \sin^{2}\theta_{13}.
\end{align}

Using equation (15) from \cite{2009PhRvD..80e3002K}, we evaluate the matter mixing angles given in their equations (16a) to (16f) in the limit where the MSW potential becomes large. Using their notation, we find that for normal mass ordering (NMO), $\tilde{\theta}_{12} \rightarrow \pi/2$, $\tilde{\theta}_{13} \rightarrow \pi/2$ for neutrinos, while for the antineutrinos ${\tilde{\theta}}_{12} \rightarrow 0$, ${\tilde{\theta}}_{13} \rightarrow 0$. The angle $\tilde{\theta}_{23}$ can be set to zero with no loss of generality if the spectra of $\mu$ and $\tau$ flavor neutrinos are taken to be equal, because any mixing between them is not observable. In the high density limit the `$\beta$' and `$\delta$' phases are also irrelevant and we can pick the Majorana phases `$\alpha_{i}$' to give positive definite values in the $U$ matrix because these phases were shown not to be observable \cite{2012JPhG...39c5201G}. Thus we find

\begin{align}
    U &= \begin{pmatrix}
     0 & 0 & 1 \\
     1 & 0 & 0 \\
     0 & 1 & 0
    \end{pmatrix},
    &
    \bar{U} &= \begin{pmatrix}
     1 & 0 & 0 \\
     0 & 1 & 0 \\
     0 & 0 & 1
    \end{pmatrix}.
    \label{eq:UsNMO}
\end{align}

For inverted mass ordering (IMO) in the same high density limit, $\tilde{\theta}_{12} \rightarrow \pi/2$, $\tilde{\theta}_{13} \rightarrow 0$ for the neutrinos, and ${\tilde{\theta}}_{12} \rightarrow 0$, ${\tilde{\theta}}_{13} \rightarrow \pi/2$ for the antineutrinos. Again, the angle ${\tilde{\theta}}_{23}$ can be set to zero with no loss of generality and again, in the high density limit, the phases are also irrelevant. This gives for the IMO

\begin{align}
    U &= \begin{pmatrix}
     0 & 1 & 0 \\
     1 & 0 & 0 \\
     0 & 0 & 1
    \end{pmatrix},
    &
    \bar{U} &= \begin{pmatrix}
     0 & 0 & 1 \\
     1 & 0 & 0 \\
     0 & 1 & 0
    \end{pmatrix}.
    \label{eq:UsIMO}
\end{align}

With $U$ and $\bar{U}$ now defined, we turn our attention to the $S$ matrices. The various prescriptions depend on the details of the flavor transformations:

\subsubsection{Adiabatic MSW Effect}
The {\bf AdiabaticMSW} transformation assumes adiabatic neutrino evolution for both mass orderings. Given the values of the mixing angle $\theta_{12}$,  $\theta_{13}$ and  $\theta_{23}$ from experiment and typical supernova density profiles, the neutrino propagation through the supernova is adiabatic in the absence of neutrino self-interactions. For adiabatic evolution, both $S_{M}$ and ${\bar{S}}_{M}$ are diagonal unitary matrices 

\begin{align}
    S_{M} &= \begin{pmatrix}
    e^{\imath\xi_{11}} & 0 & 0 \\
    0 & e^{\imath\xi_{22}} & 0 \\
    0 & 0 & e^{\imath\xi_{33}}
    \end{pmatrix},
    &
    \bar{S}_{M} &= \begin{pmatrix}
    e^{\imath{\bar{\xi}}_{11}} & 0 & 0 \\
    0 & e^{\imath{\bar{\xi}}_{22}} & 0 \\ 
    0 & 0 & e^{\imath{\bar{\xi}}_{33}}
    \end{pmatrix},
    \label{eq:Sadiabatic}
\end{align}
where the $\xi$'s and $\bar{\xi}$'s are phases which will turn out not to enter the final formulae. Assuming the NMO, this leads to
\begin{nalign}
    p_{ee} &= D_{e3}, &
    p_{ex} &= 1 - p_{ee}
    \\
    p_{xx} &= (1 + p_{ee} )/2, & 
    p_{xe} &= (1 - p_{ee} )/2
    \\
    & \\
    {\bar{p}}_{ee} &= D_{e1}, & 
    {\bar{p}}_{ex} &= 1-\bar{p}_{ee}
    \\
    {\bar{p}}_{xx} &= (1+{\bar{p}}_{ee} )/2, & 
    {\bar{p}}_{xe} &= (1-{\bar{p}}_{ee} )/2 
\end{nalign}

In the IMO, $S_{M}$ and ${\bar{S}}_{M}$ are again diagonal unitary matrices as in equations (\ref{eq:Sadiabatic}), but the final formulae are different due to the altered structure of the $U$ and $\bar{U}$ matrices at the neutrinosphere. In this mass ordering we find

\begin{nalign}
    p_{ee} &= D_{e2}, & 
    p_{ex} &= 1 - p_{ee}
    \\
    p_{xx} &= (1 + p_{ee} )/2, & 
    p_{xe} &= (1 - p_{ee} )/2
    \\
    & \\
    {\bar{p}}_{ee} &= D_{e3}, & 
    {\bar{p}}_{ex} &= 1-\bar{p}_{ee}
    \\
    {\bar{p}}_{xx} &= (1+{\bar{p}}_{ee} )/2, & 
    {\bar{p}}_{xe} &= (1-\bar{p}_{ee} )/2
\end{nalign}

\begin{figure}[t]
    \includegraphics[width=\textwidth]{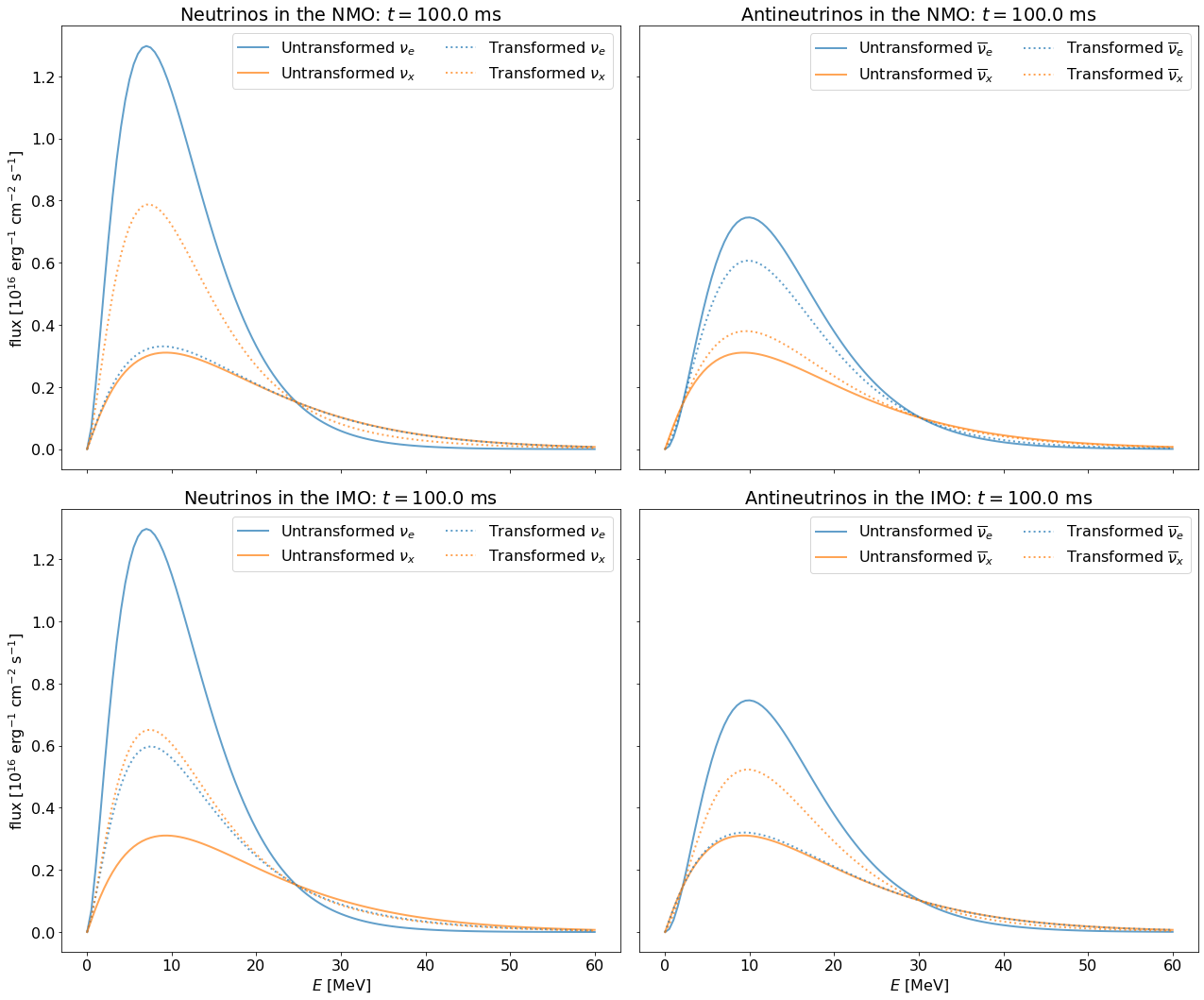}
\caption{The untransformed (solid lines) and transformed (dashed lines) spectral flux for neutrinos (left panels) and antineutrinos (right panels) using the \textbf{AdiabaticMSW} prescription. The NMO are the top pair of panels, and the IMO are the bottom  pair. We show the \textbf{nakazato-shen-z0.004-trev100ms-s20.0} model \citep{Nakazato_2013} that comes with \snewpy, at the simulation time of \SI{100}{ms} and at a supernova distance of \SI{10}{kpc}.}
\label{fig:AdiabaticMSW}
\end{figure}

In figure \ref{fig:AdiabaticMSW} we show the effect of these two prescriptions on example neutrino spectral fluxes at Earth.

\subsubsection{Nonadiabatic MSW Effect}
\begin{figure}[t]
    \includegraphics[width=\textwidth]{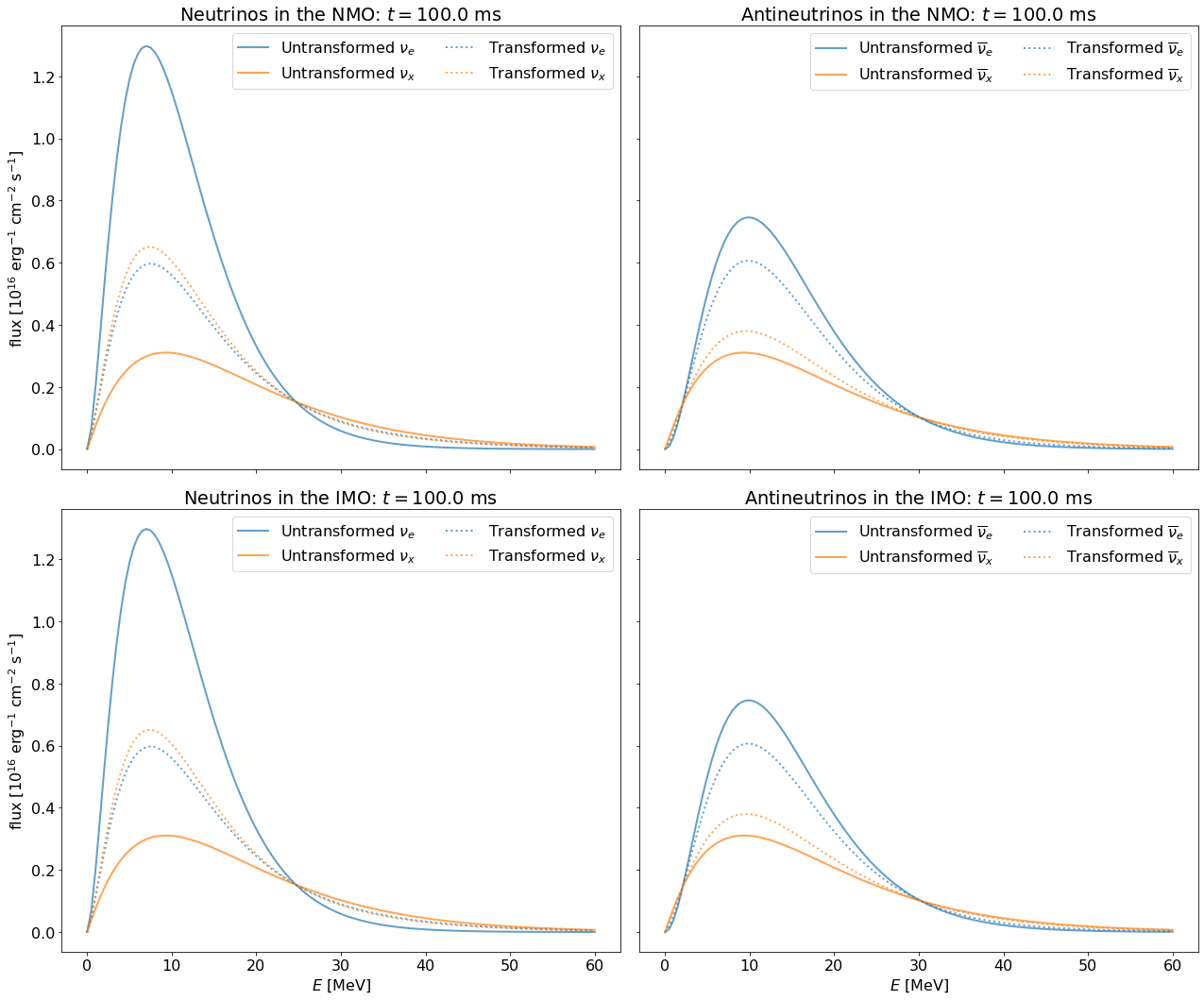}
    \caption{The same as figure \ref{fig:AdiabaticMSW} but for the \textbf{NonAdiabaticMSWH} flavor transformation prescription.}
\label{fig:NonAdiabaticMSWH}
\end{figure}

The \textbf{NonAdiabaticMSWH} transformation assumes that the H resonance mixing is nonadiabatic while the L resonance is adiabatic. This case is relevant when a shock is present at the H resonance densities \citep{2002astro.ph..5390S}. For the NMO the H resonance occurs in the neutrinos \citep{2009PhRvD..80e3002K} between `matter' states $\nu_2$ and $\nu_3$ which means the matrix $S_{M}$ is altered while ${\bar{S}}_m$ has the same diagonal structure as in equation (\ref{eq:Sadiabatic}). The new structure for $S_m$ in the NMO is

\begin{equation}
    S_{M} = \begin{pmatrix}
    e^{\imath\xi_{11}} & 0 & 0 \\ 
    0 & 0 & e^{\imath\xi_{23}} \\ 
    0 & e^{\imath\xi_{32}} & 0
    \end{pmatrix},
\end{equation}
and with this new matrix we derive
\begin{nalign}
    p_{ee} &= D_{e2}, &
    p_{ex} &= 1 - p_{ee}
    \\
    p_{xx} &= (1 + p_{ee} )/2, & 
    p_{xe} &= (1 - p_{ee} )/2
    \\
    & \\
    {\bar{p}}_{ee} &= D_{e1}, & 
    {\bar{p}}_{ex} &= 1-\bar{p}_{ee}
    \\
    {\bar{p}}_{xx} &= (1+{\bar{p}}_{ee} )/2, & 
    {\bar{p}}_{xe} &= (1-\bar{p}_{ee} )/2
\end{nalign}

In the IMO the H resonance mixes the antineutrino matter states $\bar{\nu}_1$ and $\bar{\nu}_3$. $S_{M}$ is a diagonal unitary matrix as in equation (\ref{eq:Sadiabatic}) while ${\bar{S}}_{M}$ becomes

\begin{equation}
    \bar{S}_{M} = \begin{pmatrix}
    0 & 0 & e^{\imath{\bar{\xi}}_{13}} \\ 
    0 & e^{\imath{\bar{\xi}}_{22}} & 0 \\ 
    e^{\imath{\bar{\xi}}_{31}} & 0 & 0
    \end{pmatrix}.
\end{equation}
For this case we find:
\begin{nalign}
    p_{ee} &= D_{e2}, & 
    p_{ex} &= 1 - p_{ee}
    \\
    p_{xx} &= (1 + p_{ee} )/2, &
    p_{xe} &= (1 - p_{ee} )/2
    \\
    & \\
    {\bar{p}}_{ee} &= D_{e1}, & 
    {\bar{p}}_{ex} &= 1-\bar{p}_{ee}
    \\
    {\bar{p}}_{xx} &= (1+{\bar{p}}_{ee} )/2, & 
    {\bar{p}}_{xe} &= (1-\bar{p}_{ee} )/2
\end{nalign}

In figure \ref{fig:NonAdiabaticMSWH}, we show the effect of these two prescriptions on example neutrino spectra.

\subsubsection{Two-Flavor Decoherence}
\begin{figure}[b]
    \includegraphics[width=\textwidth]{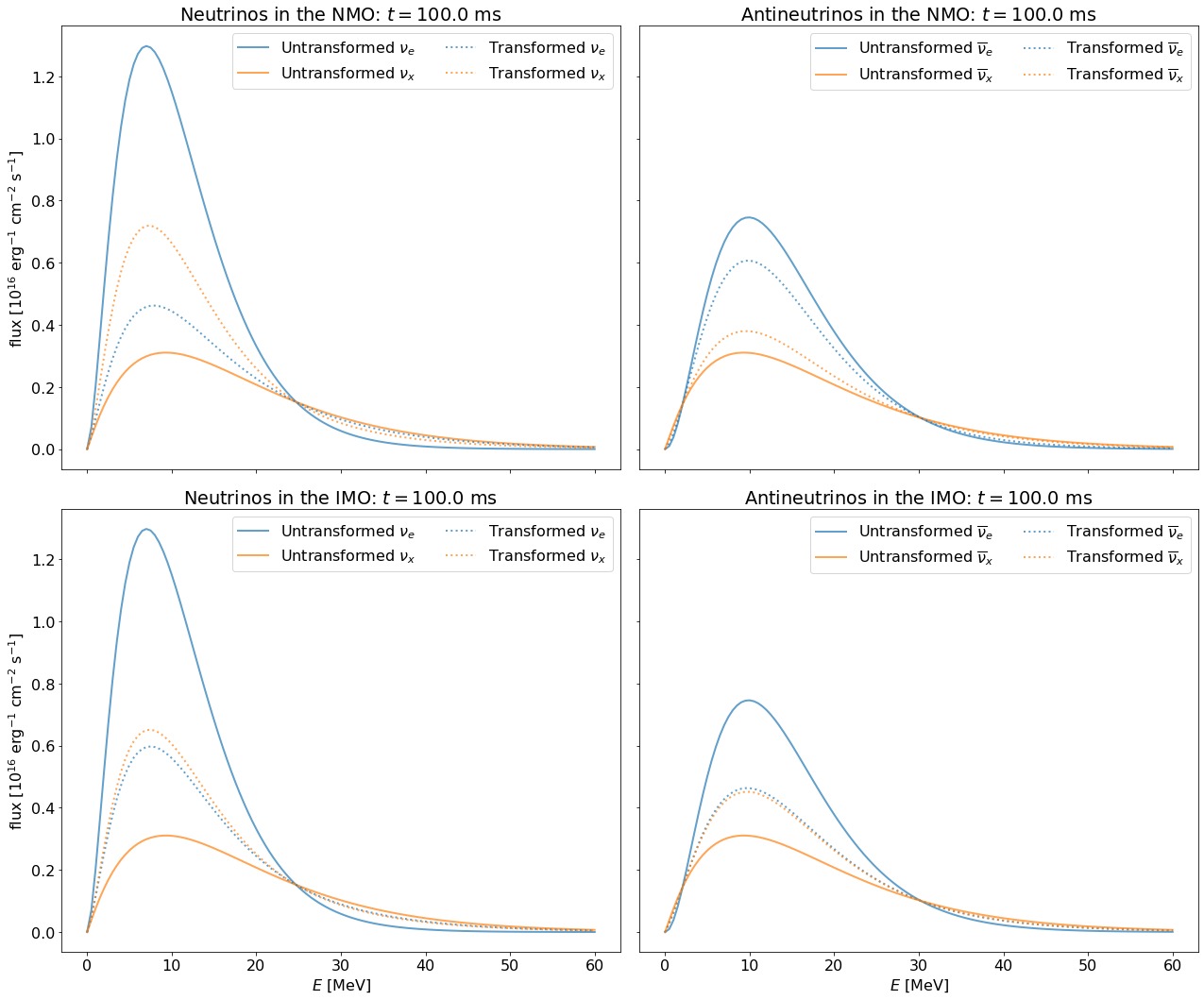}
    \caption{The same as figure \ref{fig:AdiabaticMSW} but for the \textbf{TwoFlavorDecoherence} flavor transformation prescription.}
\label{fig:TwoFlavorDecoherence}
\end{figure}

The \textbf{TwoFlavorDecoherence} transformation is relevant when there is $\lesssim 10\%$ amount of turbulence in the vicinity of the H resonance---see \cite{2010arXiv1004.1288K,2013PhRvD..88d5020K}. This prescription models 50\% mixing between the matter states which participate in the H resonance. In the NMO this is $\nu_2$ and $\nu_3$, and $S_{M}$ is

\begin{equation}
S_{M} = \frac{1}{\sqrt{2}} \begin{pmatrix}
    \sqrt{2} & 0 & 0 \\ 
    0 & e^{\imath\xi_{22}} & e^{\imath\xi_{23}} \\ 
    0 & e^{\imath\xi_{32}} & e^{\imath\xi_{33}}
    \end{pmatrix},
\end{equation}
while ${\bar{S}}_{M}$ is again a diagonal unitary matrix in this scenario. Thus we find  
\begin{nalign}
    p_{ee} &= (D_{e2}+D_{e3} ) /2 &
    p_{ex} &= 1 - p_{ee}
    \\
    p_{xx} &= (1 + p_{ee} )/2 & 
    p_{xe} &= (1 - p_{ee} )/2
    \\
    & \\
    {\bar{p}}_{ee} &= D_{e1} & 
    {\bar{p}}_{ex} &= 1-\bar{p}_{ee}
    \\
    {\bar{p}}_{xx} &= (1+{\bar{p}}_{ee} )/2 & 
    {\bar{p}}_{xe} &= (1-\bar{p}_{ee} )/2
\end{nalign}

For the IMO, the H resonance occurs in the antineutrinos between antineutrino matter states ${\bar{\nu}}_1$ and ${\bar{\nu}}_3$ so in this prescription ${\bar{S}}_M$ matrix has the general form of 

\begin{equation}
    {\bar{S}}_{M} = \frac{1}{\sqrt{2}} \begin{pmatrix}
    e^{\imath\xi_{11}} & 0 & e^{\imath\xi_{13}} \\ 
    0 & \sqrt{2} & 0 \\
    e^{\imath\xi_{31}} & 0 & e^{\imath\xi_{33}} 
    \end{pmatrix}
\end{equation}
while $S_{M}$ is diagonal. This leads to \\
\begin{nalign}
    p_{ee} &= D_{e2} & 
    p_{ex} &= 1 - p_{ee}
    \\
    p_{xx} &= (1 + p_{ee} )/2 & 
    p_{xe} &= (1 - p_{ee} )/2
    \\
    & \\
    {\bar{p}}_{ee} &= (D_{e1} + D_{e3}) /2 & 
    {\bar{p}}_{ex} &= 1-\bar{p}_{ee}
    \\
    {\bar{p}}_{xx} &= (1+{\bar{p}}_{ee} )/2 & 
    {\bar{p}}_{xe} &= (1-\bar{p}_{ee} )/2 
\end{nalign}

In figure \ref{fig:TwoFlavorDecoherence} we show the effect of these two prescriptions on example neutrino spectra

\subsubsection{Three-Flavor Decoherence}

The \textbf{ThreeFlavorDecoherence} transformation leads to 33\% mixing between all neutrino matter states and antineutrino matter states. Every element of the $S_{M}$ and ${\bar{S}}_{M}$ matrices has a magnitude of $1/\sqrt{3}$. This case is relevant when there are large amounts of turbulence in the vicinity of the H resonance no matter the mass ordering---see \cite{2013PhRvD..88d5020K}. With this structure for the $S_m$ and ${\bar{S}}_m$ matrices we obtain

\begin{nalign}
    p_{ee} &= 1/3 & 
    p_{ex} &= 1 - p_{ee}
    \\
    p_{xx} &= (1 + p_{ee} )/2 & 
    p_{xe} &= (1 - p_{ee} )/2
    \\
    & \\
    {\bar{p}}_{ee} &= 1/3 & 
    {\bar{p}}_{ex} &= 1-\bar{p}_{ee}
    \\
    {\bar{p}}_{xx} &= (1+{\bar{p}}_{ee} )/2 & 
    {\bar{p}}_{xe} &= (1-\bar{p}_{ee} )/2 
\end{nalign}

The effect of this prescription on example neutrino spectra was shown in figure~\ref{fig:ThreeFlavorDecoherence}.

\subsection{Neutrino Decay}
\snewpy also considers two cases of neutrino decay for three active neutrino flavors. The two prescriptions provided with \snewpy for neutrino decay assume adiabatic evolution through the mantle of the supernova followed by decay of the heaviest mass neutrino state to the lightest in the vacuum after they have decohered. To account for neutrino decay we must insert another matrix $G$ between the $D$ matrix and the mass state flux vector in equation (\ref{eq:F}), i.e., $D\rightarrow D\,G$. The matrix which accounts for the decay of the antineutrinos is the same. Note that we use the approximation that the energy of the neutrino does not change---future versions of \snewpy will correct this assumption.

If the heaviest neutrino has a mass $m$ and a lifetime $\tau$, we define the inverse decay length as $\Gamma = m c /( E \tau)$.
If the distance to the supernova is $d$, then the flux of the heaviest neutrino mass state decays by the factor $e^{-\Gamma d}$ and the lightest mass state increases by the amount $1+e^{-\Gamma d}$. The structure of the matrix G depends upon the mass ordering.

\begin{figure}[t]
    \includegraphics[width=\textwidth]{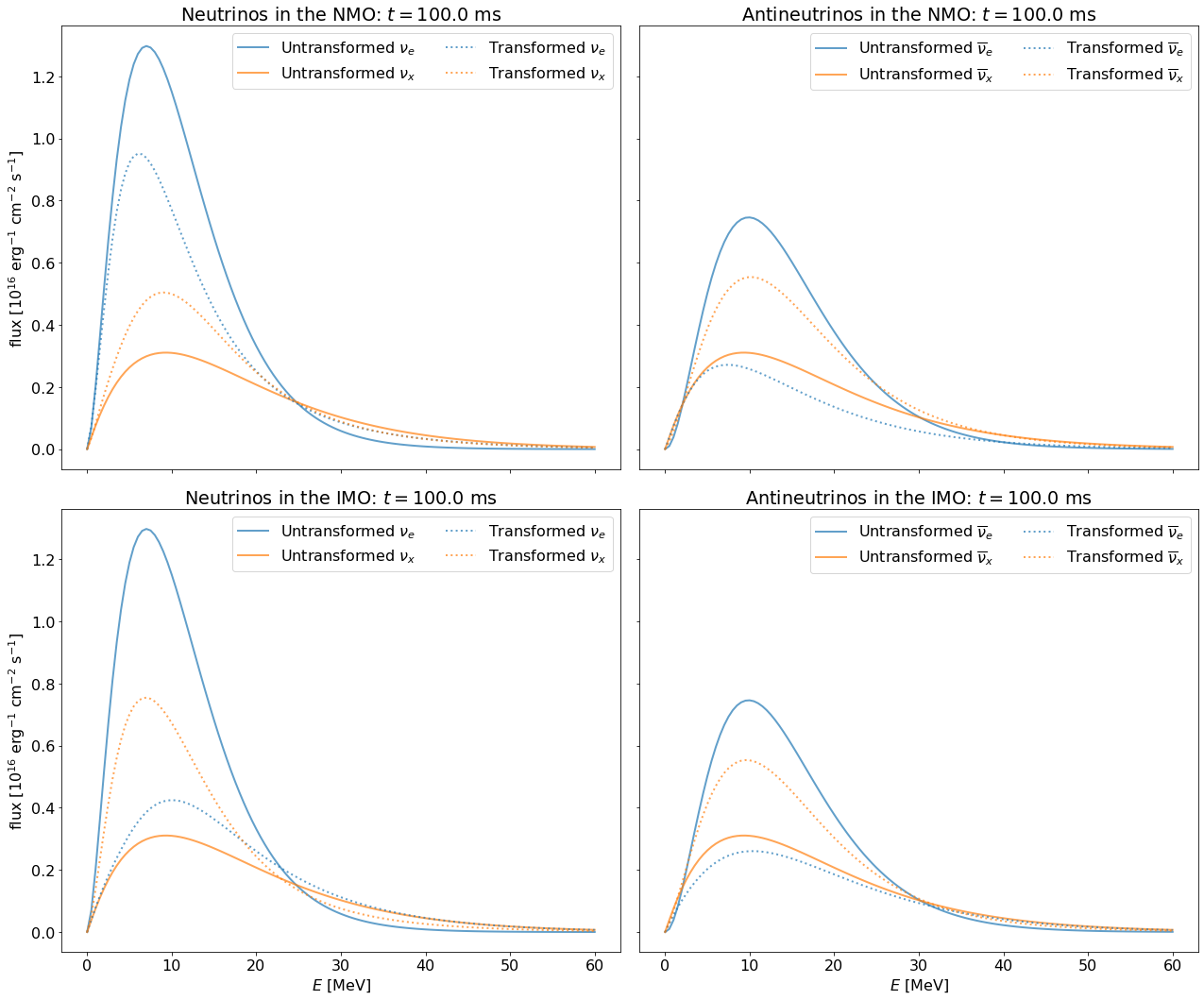}
    \caption{The same as figure \ref{fig:AdiabaticMSW} but for the \textbf{NeutrinoDecay} flavor transformation prescription. The mass of the neutrino was set to 1 eV and the mean lifetime to 1 day.}
\label{fig:NeutrinoDecay}
\end{figure}

\begin{itemize}
\item \textbf{NeutrinoDecay} -- Adiabatic evolution through the mantle of the supernova followed by neutrino decay of the heaviest mass state to the lightest without changing the neutrino energy. For the normal mass ordering the $G$ matrix is 

\begin{equation}
\setstackgap{L}{1.1\baselineskip}
\fixTABwidth{T}
    G =  \parenMatrixstack{
    1 & 0 & 1-e^{-\Gamma d} \\
    0 & 1 & 0 \\
    0 & 0 & e^{-\Gamma d}
    }
\end{equation}
which gives for the NMO:
\begin{eqnarray}
\begin{array}{ll}
p_{ee} =  D_{e1} [ 1 - e^{-\Gamma d} ] + D_{e3} e^{-\Gamma d} & \;\;\;\;\;\; p_{ex} = D_{e1} + D_{e2}\\
p_{xx} = 1 - p_{ex} /2 & \;\;\;\;\;\; p_{xe} = (1 - p_{ee} )/2\\
& \\
{\bar{p}}_{ee} = D_{e1} & \;\;\;\;\;\;
{\bar{p}}_{ex} = D_{e1} [ 1 - e^{-\Gamma d} ] +D_{e2} + D_{e3} e^{-\Gamma d}\\
{\bar{p}}_{xx} = 1-\bar{p}_{ex} /2  & \;\;\;\;\;\; {\bar{p}}_{xe} = (1-\bar{p}_{ee} )/2 
\end{array}
\end{eqnarray}

For the IMO, the $G$ matrix changes to become

\begin{equation}
\setstackgap{L}{1.1\baselineskip}
\fixTABwidth{T}
    G =  \parenMatrixstack{
    1 & 0 & 0 \\
    0 & e^{-\Gamma d} & 0 \\
    0 & 1-e^{-\Gamma d} & 1
    }
\end{equation}
and using this new matrix we find for the IMO:
\begin{eqnarray}
\begin{array}{ll}
p_{ee} = D_{e2} \exp(-\Gamma d) + D_{e3} [ 1 - \exp(-\Gamma d) ]  & \;\;\;\;\;\;
p_{ex} = D_{e1} + D_{e3}\\
p_{xx} = 1 - p_{ex} /2 & \;\;\;\;\;\; p_{xe} = (1 - p_{ee} )/2\\
& \\
{\bar{p}}_{ee} = D_{e3} & \;\;\;\;\;\;
{\bar{p}}_{ex} = D_{e1} +D_{e2} e^{-\Gamma d} + D_{e3} [ 1 - e^{-\Gamma d} ]\\
{\bar{p}}_{xx} = 1 - \bar{p}_{ex} /2   & \;\;\;\;\;\; {\bar{p}}_{xe} = (1 - \bar{p}_{ee} )/2 
\end{array}
\end{eqnarray}

In figure~\ref{fig:NeutrinoDecay}, we show the effect of this prescription on example neutrino spectra.
\end{itemize}

\subsection{The Four Neutrino Mixing Prescriptions}
We have also included prescriptions for neutrino mixing of three active flavors and a fourth, undetectable, sterile flavor. To keep things simple, we have made a number of reasonable assumptions or approximations for these prescriptions. First, we added just one more mixing angle $\theta_{14}$ and assumed the fourth mass eigenstate is the heaviest. In terms of the mixing angles, the required elements of $D$ we will require are:

\begin{align}
    D_{e1} & = \cos^2\theta_{12} \cos^2\theta_{13} \cos^2\theta_{14} \\
    D_{e2} & = \sin^2\theta_{12} \cos^2\theta_{13} \cos^2\theta_{14} \\
    D_{e3} & = \sin^2\theta_{13} \cos^2\theta_{14} \\
    D_{e4} & = \sin^2\theta_{14} \\
    D_{s1} & = \cos^2\theta_{12} \cos^2\theta_{13} \sin^2\theta_{14} \\
    D_{s2} & = \sin^2\theta_{12} \cos^2\theta_{13} \sin^2\theta_{14} \\
    D_{s3} & = \sin^2\theta_{13} \sin^2\theta_{14} \\
    D_{s4} & = \cos^2\theta_{14}.
\end{align}

In these active-sterile mixing scenarios we must consider the effect of the neutral current contribution to the MSW potential which has the effect that the initial density becomes more of an issue in determining the structure of the $U$ and $\bar{U}$ matrices. For an electron fraction $Y_e < 1/3$, the sterile flavor maps to the heaviest matter eigenstate while for $Y_e > 1/3$ it is the electron flavor which maps to the heaviest. However, in supernovae the point where $Y_e = 1/3$ is typically very close to the edge of the proto-neutron star where the density gradients are large. The current experimental limits on $\theta_{14}$ are such that it is difficult to find a supernova density profile where the $Y=1/3$ resonance is adiabatic. Thus, we have assumed that the only active-sterile mixing channel that could be either adiabatic or nonadiabatic is the $\nu_e$-$\nu_s$ resonance that is set by the mass splitting $\delta m_{14}^2$. The MSW `es' resonance is sometimes called the `outer' es resonance or $H'$ resonance \citep{1997PhRvD..56.1704N,1999PhRvC..59.2873M,2003APh....18..433F,2006PhRvD..73i3007B,2007PhRvD..76l5026K,2014PhRvD..90c3013E}. Any other new resonance introduced by adding the sterile flavor---\cite{2014PhRvD..90c3013E} calls them $H''$ resonances---is completely nonadiabatic. The mass ordering among the active flavors also needs to be specified, which gives us four cases total.

If we place the neutrinosphere between the $Y_e=1/3$ and $\nu_e$-$\nu_s$ MSW resonance then the structure of the matrices $U$ and $\bar{U}$ for the NMO are:

\begin{align}
    U &= \begin{pmatrix}
    0 & 0 & 0 & 1 \\ 
    1 & 0 & 0 & 0 \\ 
    0 & 1 & 0 & 0 \\ 
    0 & 0 & 1 & 0
    \end{pmatrix},
    &
    \bar{U} &= \begin{pmatrix} 
    1 & 0 & 0 & 0 \\ 
    0 & 0 & 0 & 1 \\ 
    0 & 0 & 1 & 0 \\ 
    0 & 1 & 0 & 0 
    \end{pmatrix}.
\end{align}

In the IMO, the matrices are:

\begin{align}
    U &= \begin{pmatrix}
    0 & 0 & 0 & 1 \\ 
    1 & 0 & 0 & 0 \\ 
    0 & 0 & 1 & 0 \\ 
    0 & 1 & 0 & 0
    \end{pmatrix}
    &
    \bar{U} &= \begin{pmatrix}
    0 & 0 & 1 & 0 \\ 
    0 & 1 & 0 & 0 \\ 
    0 & 0 & 0 & 1 \\ 
    1 & 0 & 0 & 0 
    \end{pmatrix}.
\end{align}

Putting this all together we find the following formulae in the four cases we consider.

\begin{figure}[b]
    \includegraphics[width=\textwidth]{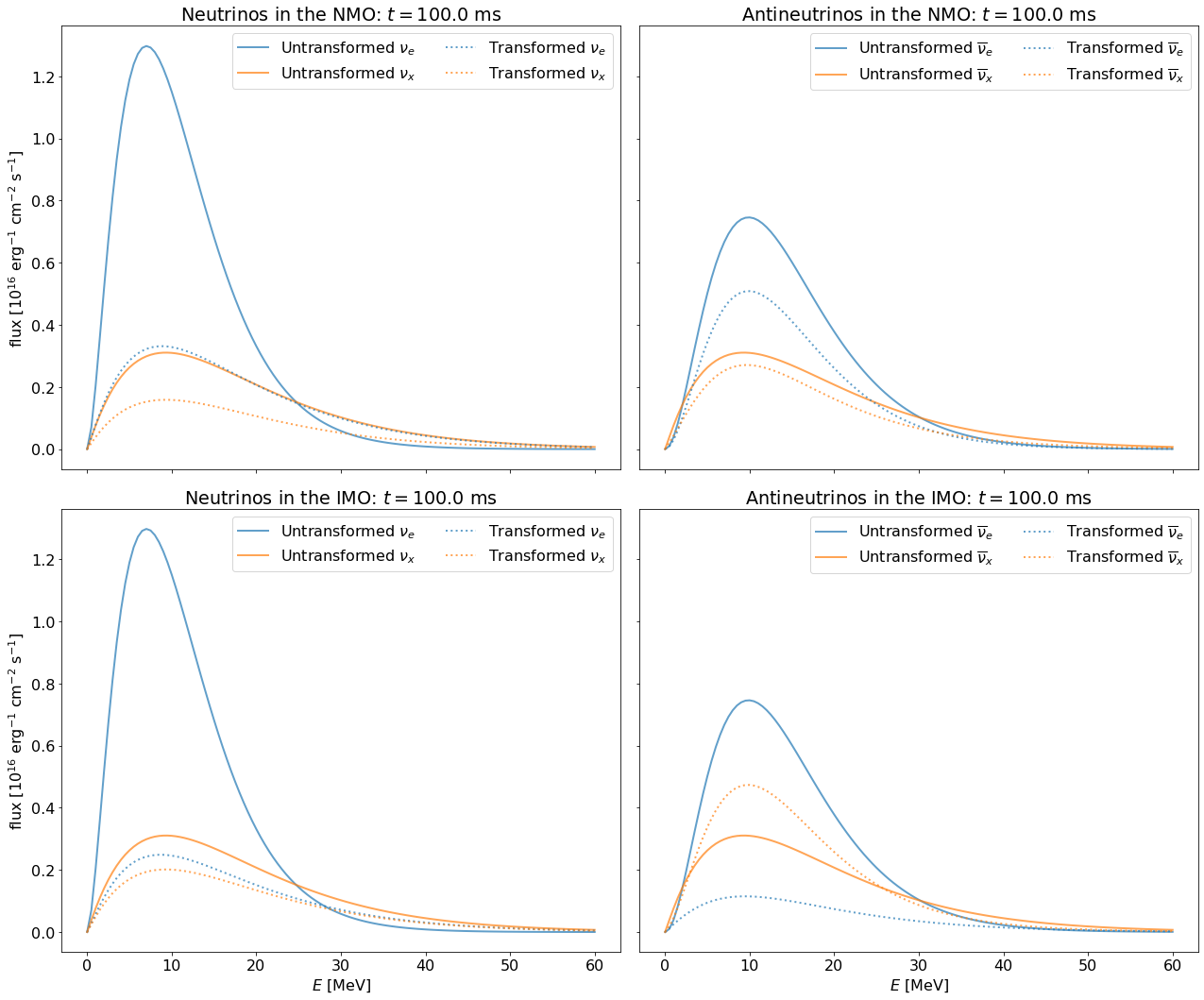}
    \caption{The same as figure \ref{fig:AdiabaticMSW} but for the \textbf{AdiabaticMSWes} flavor transformation prescription. The mixing angles $\theta_{12}$, $\theta_{13}$ and $\theta_{23}$ are set to the values from \cite{Zyla:2020zbs}, the mixing angle $\theta_{14} = 10^{\circ}$. }
\label{fig:AdiabaticMSWes}
\end{figure}

\begin{itemize}
\item \textbf{AdiabaticMSWes}: For the NMO:

\begin{nalign}
    p_{ee} &= D_{e4},  & & &  
    p_{ex} &= D_{e1} + D_{e2}
    \\
    p_{xx} &= ( 2 - D_{e1} - D_{e2} - D_{s1} - D_{s2} ) / 2,  & & &
    p_{xe} &= (1-D_{e4}-D_{s4} )/2 \\
    & \\
    {\bar{p}}_{ee} &= D_{e1}, & & &
    {\bar{p}}_{ex} &= D_{e3} + D_{e4}
    \\
    {\bar{p}}_{xx} &= ( 2 - D_{e3} - D_{e4} - D_{s3} - D_{s4}) / 2, & & &
    {\bar{p}}_{xe} &= ( 1 - D_{e1} - D_{s1} ) / 2
\end{nalign}

For the IMO:

\begin{nalign}
    p_{ee} &= D_{e4}, & & & 
    p_{ex} &= D_{e1} + D_{e3}
    \\
    p_{xx} &= ( 2 - D_{e1} - D_{e3} - D_{s1} - D_{s3} ) / 2, & & &
    p_{xe} &= ( 1 - D_{e4}  - D_{s4} ) / 2
    \\
    & \\
    p_{ee} &= D_{e4}, & & &
    p_{ex} &= D_{e1} + D_{e3}
    \\
    p_{xx} &= ( 2 - D_{e1} - D_{e3} - D_{s1} - D_{s3} ) / 2, & & &
    p_{xe} &= ( 1 - D_{e4}  - D_{s4} ) / 2
\end{nalign}

In figure~\ref{fig:AdiabaticMSWes}, we show the effect of these prescriptions on example neutrino spectra.

\begin{figure}[b]
    \includegraphics[width=\textwidth]{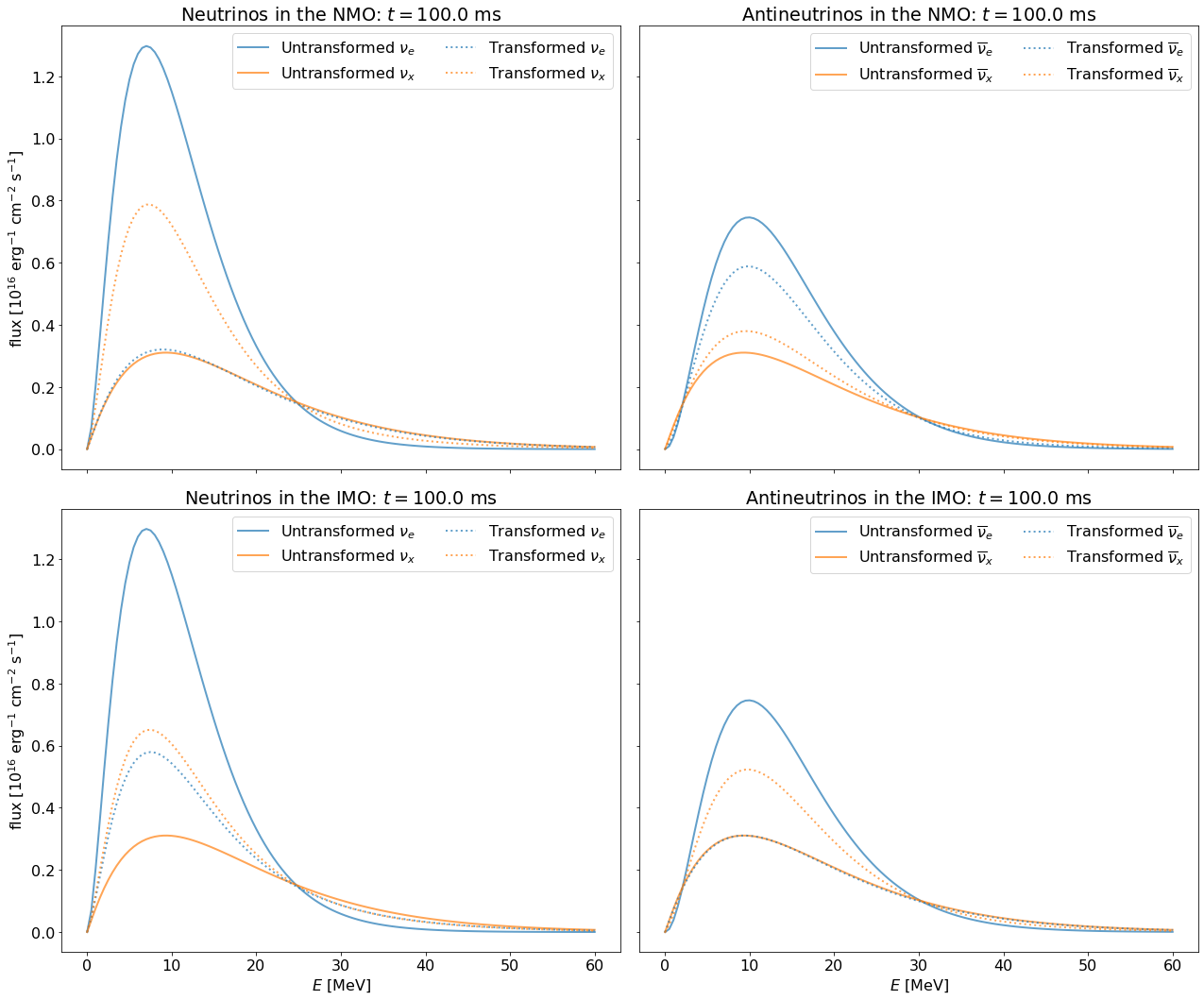}
    \caption{The same as figure \ref{fig:AdiabaticMSW} but for the \textbf{NonAdiabaticMSWes} flavor transformation prescription. The mixing angles $\theta_{12}$, $\theta_{13}$ and $\theta_{23}$ are set to the values from \cite{Zyla:2020zbs}, the mixing angle $\theta_{14} = 10^{\circ}$. }
\label{fig:NonAdiabaticMSWes}
\end{figure}

\item \textbf{NonAdiabaticMSWes}: For the NMO, the $\nu_e$-$\nu_s$ resonances in the neutrinos, and the two resonances in the antineutrinos---see the eigenvalue diagram in \cite{2014PhRvD..90c3013E}---are all nonadiabatic. The structure of the $S$ and $\bar{S}$ matrices are thus:

\begin{align}
    S_{M} &= \begin{pmatrix}
    e^{\imath\xi_{11}} & 0 & 0 & 0 \\ 
    0 & e^{\imath\xi_{22}} & 0 & 0 \\ 
    0 & 0 & 0 & e^{\imath\xi_{34}} \\ 
    0 & 0 & e^{\imath\xi_{43}} & 0 
    \end{pmatrix}
    &
    {\bar{S}}_{M} &= \begin{pmatrix}
    e^{\imath\xi_{11}} & 0 & 0 & 0 \\ 
    0 & 0 & e^{\imath\xi_{23}} & 0 \\ 
    0 & 0 & 0 & e^{\imath\xi_{34}} \\ 
    0 & e^{\imath\xi_{42}} & 0 & 0
    \end{pmatrix}.
\end{align}
which interestingly leads to the following formulae:
\begin{nalign}
    p_{ee} &= D_{e3}, & & &
    p_{ex} &= D_{e1} + D_{e2}
    \\
    p_{xx} &= ( 2 - D_{e1} - D_{e2} - D_{s1} - D_{s2} ) / 2, & & &
    p_{xe} &= (1-D_{e3}-D_{s3} ) / 2
    \\
    & \\
    {\bar{p}}_{ee} &= D_{e1}, & & &
    {\bar{p}}_{ex} &= D_{e2} + D_{e3}
    \\
    {\bar{p}}_{xx} &= ( 2 - D_{e2} - D_{e3} - D_{s2} - D_{s3} ) / 2, & & &
    {\bar{p}}_{xe} &= ( 1 - D_{e1} - D_{s1} ) / 2.
\end{nalign}

For the IMO the $\nu_e$-$\nu_s$ resonance in the neutrinos and the two resonances in the antineutrinos are all non-adiabatic, and the mass ordering of the active flavors is inverted. The $S$ and $\bar{S}$ matrices are

\begin{align}
S_{M}  &=  \begin{pmatrix}
    e^{\imath\xi_{11}} & 0 & 0 & 0 \\
    0 & 0 & 0 & e^{\imath\xi_{24}} \\ 
    0 & 0 & e^{\imath\xi_{33}} & 0 \\ 
    0 & e^{\imath\xi_{42}} & 0 & 0
    \end{pmatrix}
    &
{\bar{S}}_{M} &= \begin{pmatrix}
    0 & e^{\imath\xi_{12}} & 0 & 0 \\ 
    0 & 0 & 0 & e^{\imath\xi_{24}} \\ 
    0 & 0 & e^{\imath\xi_{33}} & 0 \\ 
    e^{\imath\xi_{41}} & 0 & 0 & 0
    \end{pmatrix},
\end{align}
which leads to
\begin{nalign}
    p_{ee} &= D_{e2}, & & & 
    p_{ex} &= D_{e1} + D_{e3}\\
    p_{xx} &= ( 2 - D_{e1} - D_{e3} - D_{s1} - D_{s3} ) / 2, & & &
    p_{xe} &= ( 1 - D_{e2} - D_{s2} ) / 2\\
    & \\
    {\bar{p}}_{ee} &= D_{e3}, & & & 
    {\bar{p}}_{ex} &= D_{e1} + D_{e2}\\
    {\bar{p}}_{xx} &= ( 2 - D_{e1} - D_{e2} - D_{s1} - D_{s2} ) / 2, & & & 
    {\bar{p}}_{xe} &= ( 1 - D_{e3} - D_{s3} ) / 2
\end{nalign}
for both mass orderings.
In figure~\ref{fig:NonAdiabaticMSWes}, we show the effect of these prescriptions on example neutrino spectra.

\end{itemize}

\bibliography{references}


\end{document}